\documentclass[a4paper,fleqn,usenatbib]{mnras}

\usepackage{graphicx} 
\usepackage{txfonts} 
\usepackage{natbib} 
 
\newcommand{\Mpc}{h^{-1}\, {\rm Mpc}}

\newcommand{\be}{\begin{equation}}
\newcommand{\ee}{\end{equation}}
\newcommand{\bea}{\begin{equation}\begin{aligned}} 
\newcommand{\eea}{\end{aligned}\end{equation}}

\def\apj{ApJ} 
\def\apjl{ApJL} 
\def\apjs{ApJS} 
 
\def\aap{A\&A} 
\def\mnras{MNRAS}

 \title{The time evolution of bias}
 
\author[J. Einasto et al.]{
J. Einasto,$^{1,2,3}$\thanks{E-mail: jaan.einasto@ut.ee} 
 L. J. Liivam\"agi,$^{1}$
and M. Einasto$^{1}$
\\
$^{1}$Tartu Observatory, 61602 T\~oravere, Estonia\\  
$^{2}$ICRANet, Piazza della Repubblica 10, 65122 Pescara, Italy \\ 
$^{3}$Estonian Academy of Sciences, 10130 Tallinn, Estonia\\
} 

\date{Accepted XXX. Received YYY; in original form ZZZ}
\pubyear{2022}

\begin{document}
\label{firstpage}
\pagerange{\pageref{firstpage}--\pageref{lastpage}}
\maketitle
 
\begin{abstract}
  
  We investigate the time evolution of bias of cosmic density fields.
  We perform numerical simulations of the evolution of the cosmic web
  for the conventional $\Lambda$ cold dark matter ($\Lambda$CDM)
  model. The simulations cover a wide range of box sizes
  $L=256 - 1024\Mpc$, and epochs from very early moments $z=30$ to the
  present moment $z=0$.  We calculate spatial correlation functions of
  galaxies, $\xi(r)$, using dark matter particles of the biased
  $\Lambda$CDM simulation.  We analyse how these functions describe
  biasing properties of the evolving cosmic web. We find that for all
  cosmic epochs the bias parameter, defined through the ratio of
  correlation functions of selected samples and matter, depends on two
  factors: the fraction of matter in voids and in the clustered
  population, and the luminosity (mass) of galaxy samples. Gravity
  cannot evacuate voids completely, thus there is always some
  unclustered matter in voids, thus the bias parameter of galaxies is
  always greater than unity, over the whole range of evolution
  epochs. We find that for all cosmic epochs bias parameter values
  form regular sequences, depending on galaxy luminosity (particle
  density limit), and decreasing with time.
  
\end{abstract}


\section{Introduction} 

The relative distribution of galaxies and mass and its evolution in
time is of increasing concern in cosmology. It is well known that
galaxies do not trace exactly the mass.  The difference in the
distribution of mass and galaxies was noticed already  by
\citet{ Joeveer:1978dz, Joeveer:1978pb}, who found in the distribution
of galaxies  large almost empty voids, occupying about 98 per cent of
the volume of the universe. Authors concluded that since gravity works
slowly it is impossible to evacuate such large regions completely --
there must be dark matter (DM) in voids. The presence of rarefied
matter in voids  was demonstrated  by early numerical
simulations of the evolution of the universe by
\citet{Doroshkevich:1980, Doroshkevich:1982fk}. The difference between
  distributions of matter and galaxies was explained by
\citet{Zeldovich:1982kl} as an indication of a treshold mechanism in
galaxy formation -- in low-density regions galaxies cannot form. This
phenomenon was described quantitatively by \citet{Kaiser:1984}, who
suggested the term ``biasing''. In this paper \citet{Kaiser:1984}
used biasing to describe  the difference in the distribution of
galaxies and clusters of galaxies. Novadays this term denotes the  
relationship between  distributions of galaxies of various
luminosity (or mass) and that of the mass, including DM.

There exist a very large number of studies devoted to the biasing
problem, for a recent review see \citet{Desjacques:2018qf}.  Most of
these studies discuss detailed properties of the distribution of
galaxies and matter in the present epoch.  There exist only a few
studies of the time evolution of the bias. These studies are based
either on the perturbation theory of the evolution of the cosmic web,
as done by \citet{Fry:1996vj} and \citet{Tegmark:1998yq}, or on
numerical simulation of the evolution of the density field
(\citet{Dubois:2021tq}, \citet{Park:2022va}).

Usually  the bias is defined through  density fields of matter and
galaxies: $b = \delta_g/\delta_m$, where
$\delta_g=N_{\mathrm{gal}}/\overline{N} - 1$ and
$\delta_m=\rho_m/\overline{\rho}-1$ are density contrasts of galaxies
and matter \citep{Desjacques:2018qf}.  However, as noted by
\citet{Repp:2019wl, Repp:2019ti, Repp:2020vr, Repp:2020wx}, this bias
model leads to unphysical results, since in voids the galaxy density
is zero and density contrast is negative, $\delta_g =-1$.  To avoid
this defect we define the bias function through  correlation functions
of galaxies and matter, $b^2 =\xi_g/\xi_m$, as done in the
pioneering work by \citet{Kaiser:1984}.

So far in bias studies only a few cosmic epochs were considered.  To
understand the bias phenomenon in a broader cosmological context it is
needed to find the relationship between matter and galaxies over a
large interval of cosmic epochs.  The goal of this study is to
investigate the time evolution of bias in a large time interval using
numerical simulations of the evolution of the cosmic density field.
We assume that the $\Lambda$ cold dark matter ($\Lambda$CDM) model
represents the actual universe accurately enough, and that it can be
used to investigate the evolution of the structure of the real
universe.  To study the time evolution of bias we use a set of
numerical simulations with number of particles
$N_{\mathrm{part}} =512^3$ and simulation cube sizes
$L_0=256,~512,~1024~\Mpc$. These simulations were used earlier by
\citet{Einasto:2019aa, Einasto:2020aa, Einasto:2021uz, Einasto:2021ti}
to investigate various properties of the cosmic web.

The bias analysis with correlation functions can be done using three
types of objects: simulation particles, simulated galaxies and 
real galaxies. The analysis by \citet{Einasto:2019aa, Einasto:2020aa} has shown
that all three types of objects yield for the bias function similar
results.  In the present paper we shall use simulation particles as
test objects, this approach gives most accurate correlation functions
for a wide interval of particle separations.  As done by
\citet{Einasto:2019aa}, we use a simple biased DM simulation model, and
divide matter into a low-density population with no galaxy formation
or a population of simulated galaxies below a certain mass limit, and
a high-density population of  clustered matter, associated with
galaxies above the mass limit.  For each simulation epoch we use
(Eulerian) particle local densities, $\rho$, and label particles with
this density value.

We apply a sharp particle-density limit, $\rho_0$, to select biased
samples of particles.  This method to
select biased galaxy (particle) models was applied earlier among
others by \citet{Jensen:1986aa}, \citet{Einasto:1991fq},
\citet{Szapudi:1993aa}, and \citet{Little:1994rt}. This model to
select particles for simulated galaxies is similar to the Ising model,
discussed by \citet{Repp:2019ti, Repp:2019wl}. Actually galaxy
formation is a stochastic process, thus the matter density limit which
divides unclustered and clustered matter is fuzzy
(\citet{Dekel:1998wz}, \citet{Dekel:1999wn},
\citet{Taruya:1999uw,Taruya:1999vq}, \citet{Tegmark:1999vn}).
However, a fuzzy density limit has little influence on properties of
correlation functions or power spectra of biased and non-biased
samples, as shown by \citet{Einasto:2019aa}.  Biased model samples
include particles with density labels, $\rho \ge \rho_0$.  These
samples are found from the full DM sample by excluding particles of
density labels less than the limit $\rho_0$.  In this way biased model
samples mimic observed samples of galaxies, where there are no
galaxies fainter than a certain luminosity limit.  We use a series of
particle-density limits $\rho_0$.

The paper is organised as follows. In the following Section we
describe numerical simulations used in our study, the calculation of
the density fields of simulated samples, and the method
to find correlation functions. In Section 3 we describe the evolution
of particle densities and of the density field. In Section 4 we
describe the evolution of biasing properties of our simulated
universe. Sections 5 and 6 present the discussion of results and
conclusions.

{\scriptsize 
\begin{table}
\caption{Parameters of simulations }
\centering
\begin{tabular}{lrrrrc}
\hline  \hline
  Simulation   & $L_0$&$\sigma_8$&$m_p$&$R_0$\\   
\hline  
(1)&(2)&(3)&(4) &(5)\\
  \hline  \\
  L256& 256&0.613&$0.993$&0.5\\
  L512& 512&0.641&$7.944$& 1.0\\
  L1024&1024&0.646&$63.55$& 2.0\\  
 \label{Tab1}                         
\end{tabular} \\
{Columns give: (1) name of simulation; (2) box size in
  $\Mpc$; (3) $\sigma_8$ (4)
  mass of a particle in units of $10^{10}h^{-1}M_\odot$; and (5)
  effective smoothing scale $R_0$ in units of $\Mpc$.} 
\end{table} 
}

\begin{figure*}
\centering 
\hspace{2mm}
\resizebox{0.35\textwidth}{!}{\includegraphics*{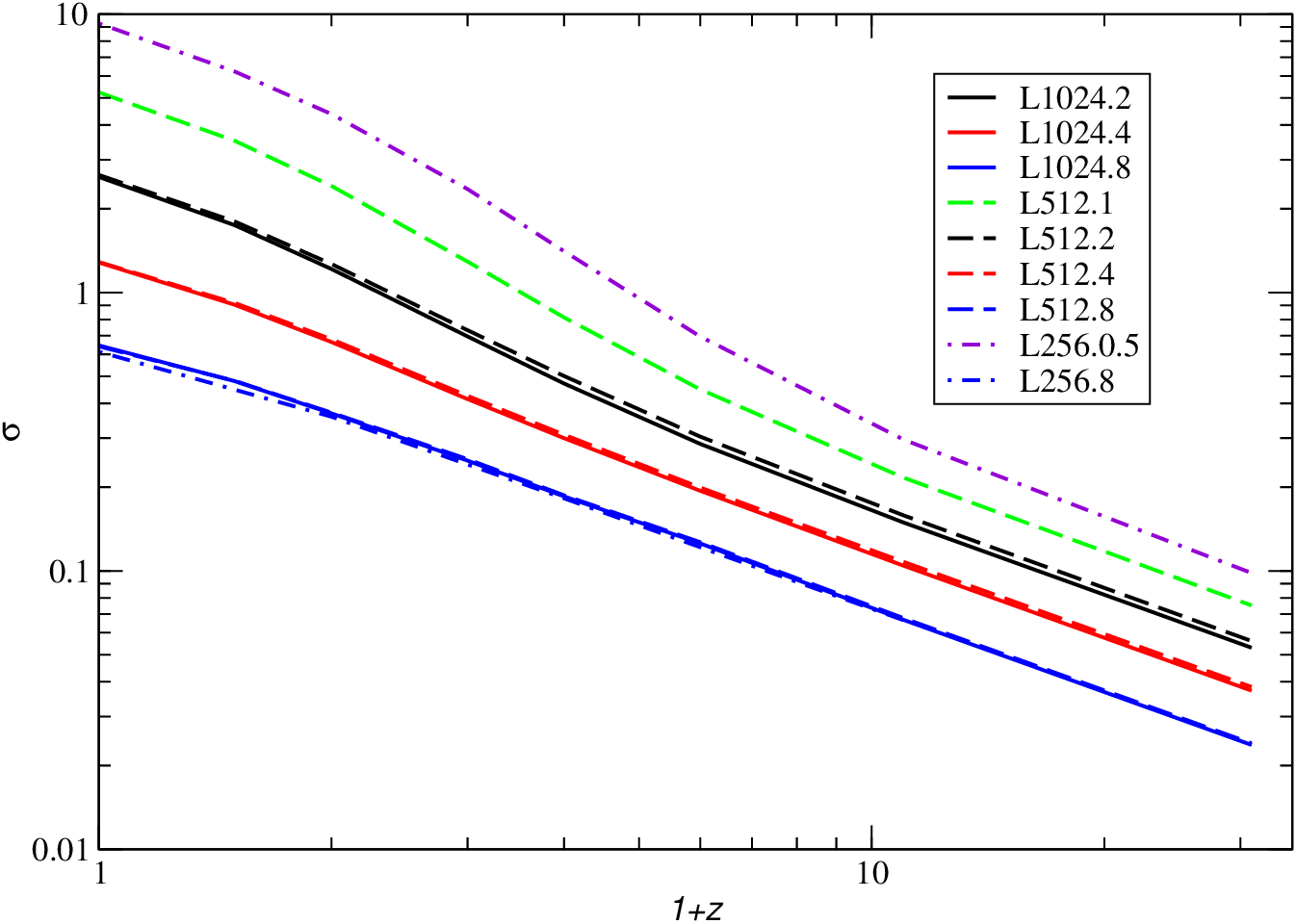}}
\hspace{2mm}
\resizebox{0.35\textwidth}{!}{\includegraphics*{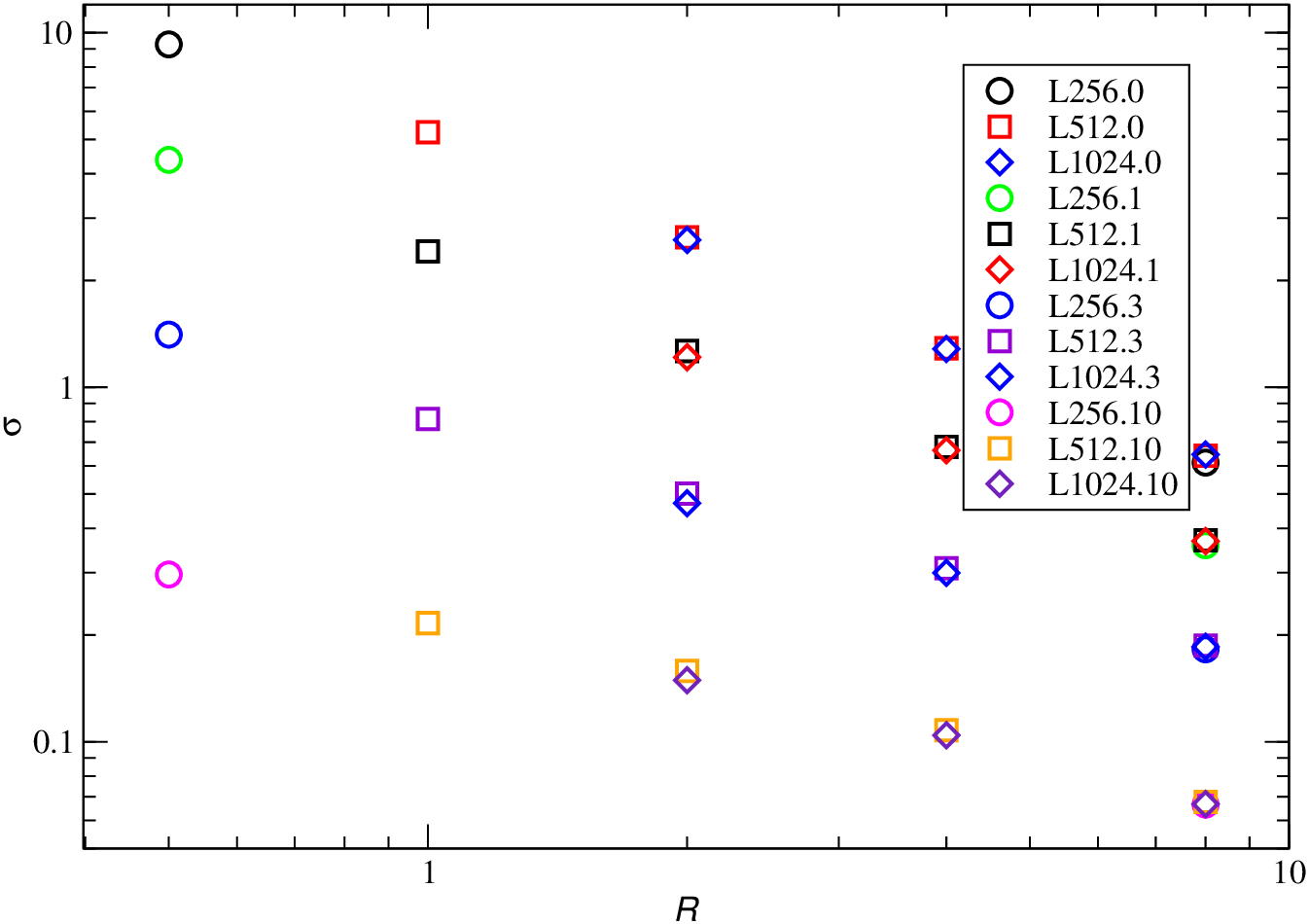}}
\caption{Left: Dependence of the dispersion $\sigma$ of models L256,
  L512, and L1024 on simulation epoch $z$.  The second index in model
  name shows the smoothing scale $R$ in $\Mpc$. Line types are for
  models of different size: solid lines for L1024, dashed
  lines for L512, dotted and dot-dashed lines for L256; colours show
  smoothing scale: blue for $R=8~\Mpc$, red for $R=4~\Mpc$, black for
  $R=2~\Mpc$, green for $R=1~\Mpc$, and violet for $R=0.5~\Mpc$.
  Right: Dependence of the dispersion $\sigma$ on the smoothing length
  $R$ of the same models.  Here the second index of the model name
  gives the redshift $z$. The upper sequence of symbols is for the
  present epoch $z=0$, lower sequences are for redshifts 
  $z=1,~3,~10$, respectively. }
\label{fig:Fig1} 
\end{figure*} 

\section{Data and methods}

In this Section we describe the calculation of simulations and the
finding of particle density limited samples, which can be called also
as simulated galaxy samples.  Then we describe the calculation of density
fields and the calculation of correlation and bias functions. We use
the value of bias functions at separation $r_6=6$ and $r_{10}=10$ as bias
parameters.

\subsection{Calculation of simulations  and biased  model  samples}

We simulated the evolution of the cosmic web adopting a DM
$\Lambda$CDM model.  We use the {\sc GADGET} code
\citep{Springel:2005} with three different box sizes
$L_0=256,~512,~1024~\Mpc$ with $N_{\mathrm{grid}} = 512$, and number
of particles $N_{\mathrm{part}} = 512^3$.  We call these simulations
as L256, L512, and L1024 models.  Cosmological parameters for all 
simulations are ($\Omega_m,\Omega_{\Lambda},\Omega_b,h,\sigma_8,n_s$)
=(0.28,~0.72,~0.044,~0.693,~0.84,~1.00).  Initial conditions were
generated using the COSMICS code by \citet{Bertschinger:1995},
assuming Gaussian fluctuations.  Simulations started at redshift
$z=30$ using the Zeldovich approximation. We extracted density fields
and particle coordinates for eight epochs, corresponding to redshifts
$z=30,~10,~5,~3,~2,~1,~0.5,~0$.  Table \ref{Tab1} shows parameters of
simulations.

For all simulation particles and all simulation epochs, we calculated
the local density values at particle locations, $\rho$, using the
positions of the 27 nearest particles, including the particle itself.
Densities were expressed in units of the mean density of the whole
simulation.  The full $\Lambda$CDM model includes all particles.
Following \citet{Einasto:2019aa}, we formed biased model samples that
contained particles above a certain limit, $\rho \ge \rho_0$, in units
of the mean density of the simulation.  As shown by
\citet{Einasto:2019aa}, this sharp density limit allows to find
density fields of simulated galaxies, whose geometrical properties are
close to the density fields of real galaxies.  Thus biased model
samples can be called also as simulated galaxy samples or as clustered
samples, depending on the context.  This method to select particles
for simulated galaxies is similar to the Ising model, discussed by
\citet{Repp:2019ti, Repp:2019wl}.  We use a series of particle-density
limits $\rho_0$ for each simulation epoch.  Biased samples are found
from the full DM sample by excluding particles with density labels
less than the limit $\rho_0$.  In this way biased model samples mimic
observed samples of galaxies, where there are no galaxies fainter than
a certain luminosity limit.

The biased samples are denoted LXXX.$i$, where  XXX denotes the size
of the simulation box in $\Mpc$, and $i$ denotes the
particle-density limit $\rho_0$.  The full DM model includes all
particles and corresponds to the particle-density limit
$\rho_0 = 0$,  therefore it is denoted as LXXX.00.  The main data
on the model samples are given in Table~\ref{Tab1}.

\subsection{Calculation of density fields}

$N-$body simulations provides us with  populations of DM particles  in a box of
size $L_0$ at redshift $z$. The density field was estimated using a
filter of size $R_t$.  The density field was normalised to the average
matter density, providing us with the relative  density  $D(\mathbf{x})$,
\be
  D(\mathbf{x})= \frac{\rho(\mathbf{x})}{\overline{\rho}},
\ee
where  $D(\mathbf{x})$ is the density at location $\mathbf{x}$, and 
${\overline{\rho}}$ is the mean density. 
The second moment of  the density contrast  $\delta(\mathbf{x}) =
D(\mathbf{x}) - 1$  is the variance of the density field,
$\sigma^2$. In the following we call $\sigma$ as the dispersion of the
density field. 

We determined  smoothed density fields of simulations
using a $B_3$ spline \citep[see][]{Martinez:2002fu},
\begin{equation} 
B_3(x)=\frac1{12}\left[|x-2|^3-4|x-1|^3+6|x|^3-4|x+1|^3+|x+2|^3\right]. 
\end{equation}
The spline function is different from zero only in the interval
$x\in[-2,2]$.  To calculate the high-resolution density field we used
the kernel of the scale, which is equal to the cell size of the
simulation, $L_0/N_{\mathrm{grid}}$, where $L_0$ is the size of the
simulation box and $N_{\mathrm{grid}}$ is the number of grid elements
in one coordinate.  The smoothing with index $i$ has a smoothing
radius $R_i= L_0/N_{\mathrm{grid}} \times 2^i$. The effective scale of
smoothing is equal to $2\times R_i$.  Density fields extracted from
simulations have effective smoothing scale
$R_0= L_0/N_{\mathrm{grid}}$. We calculated density fields of models
up to smoothing scale $R_i=8~\Mpc$. The $B_3$ kernel of radius
$R_B=1~\Mpc$ corresponds to a Gaussian kernel with dispersion
$R_G = 0.6~\Mpc$.  For details of the smoothing method see Appendix.A
of \citet{Einasto:2021uz}.

\subsection{Calculation of  the correlation and bias functions}

The $\Lambda$CDM model samples contain all particles with local
density labels $\rho \ge \rho_0$.  To derive correlation functions of
these samples, a conventional methods \citep{Landy:1993ve} cannot be
used because the number of particles is too large, up to $512^3$ in
the full unbiased model.  To determine the correlation functions we
used the \citet{Szapudi:2005aa} method. 
This method uses the FFT to calculate correlation functions and scales
as $O(N \log N)$.  The method is an implementation of the algorithm
{\tt eSpICE}, the Euclidean version of {\tt SpICE} by
\citet{Szapudi:2001aa}.  To find the dependence of results on the size
of the grid we calculated correlation functions with two sizes of the
grid, $N_{\mathrm{grid}}=2048^3$ and $N_{\mathrm{grid}}=1024^3$.  The
finer grid allows to see better the shape of correlation functions and
its derivatives on halo scales, the coarse grid allows to suppress
wiggles of the correlation functions on large separations.  For the
finer grid correlation functions were found up to separations
$r_{max}=204~\Mpc$ to $r_{max}=290~\Mpc$ with 308 logarithmic bins for
models L256 to L1024. For the $N_{\mathrm{grid}}=1024^3$ grid we calculated correlation
functions up to separation $r_{max}=64~\Mpc$, $r_{max}=127~\Mpc$ and
$r_{max}=255~\Mpc$ for models L256, L512 and L1024, using 128 linear
bins.  The analysis of both sets of correlation and bias functions
showed that wiggles of both functions found with the
$N_{\mathrm{grid}}=2048^3$ grid are too large for reliable
determination of bias parameters, thus in the following we use only
results obtained with the $N_{\mathrm{grid}}=1024^3$ grid.

\begin{figure*}
\centering 
\hspace{1mm}
\resizebox{0.32\textwidth}{!}{\includegraphics*{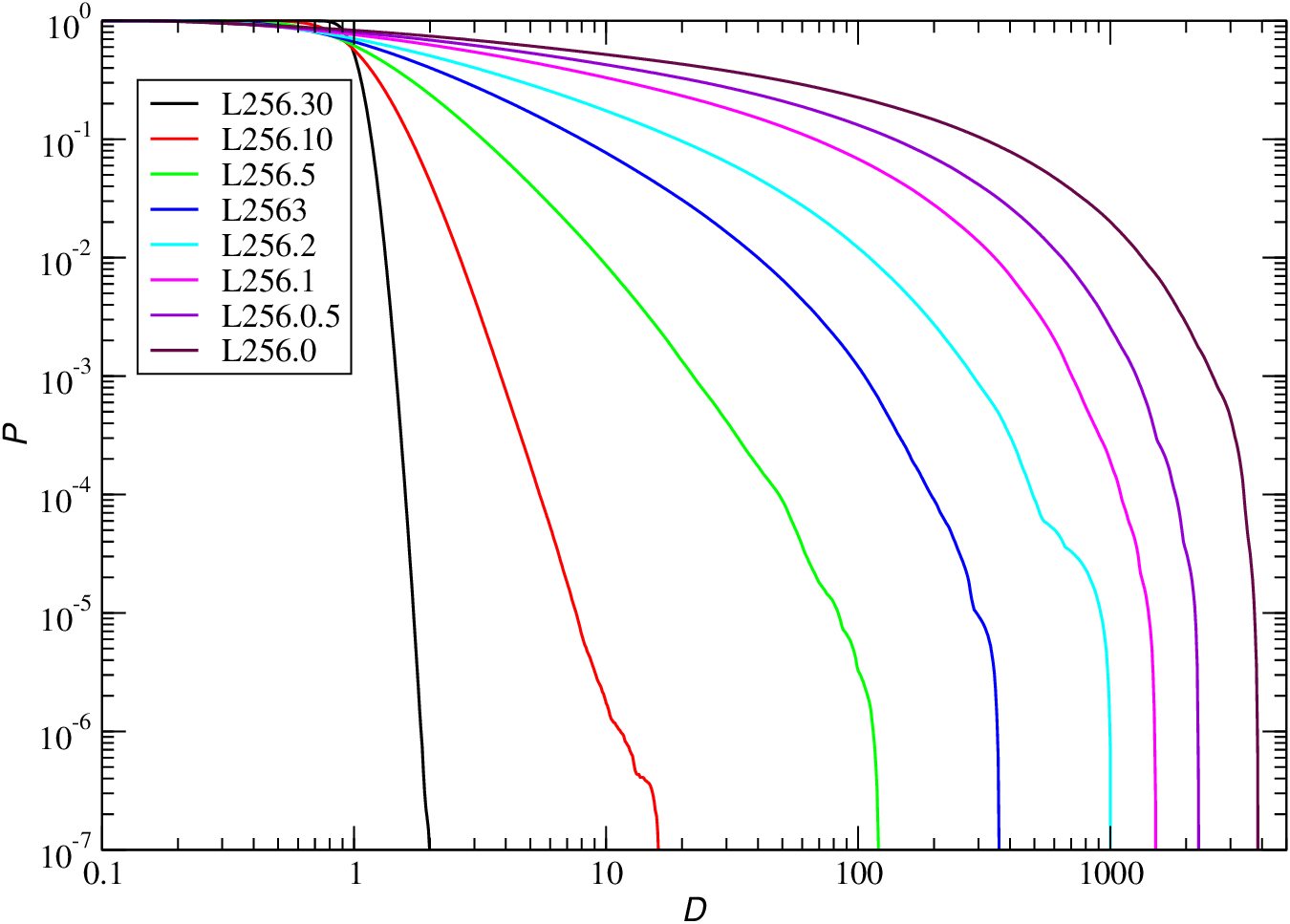}}
\hspace{1mm}
\resizebox{0.32\textwidth}{!}{\includegraphics*{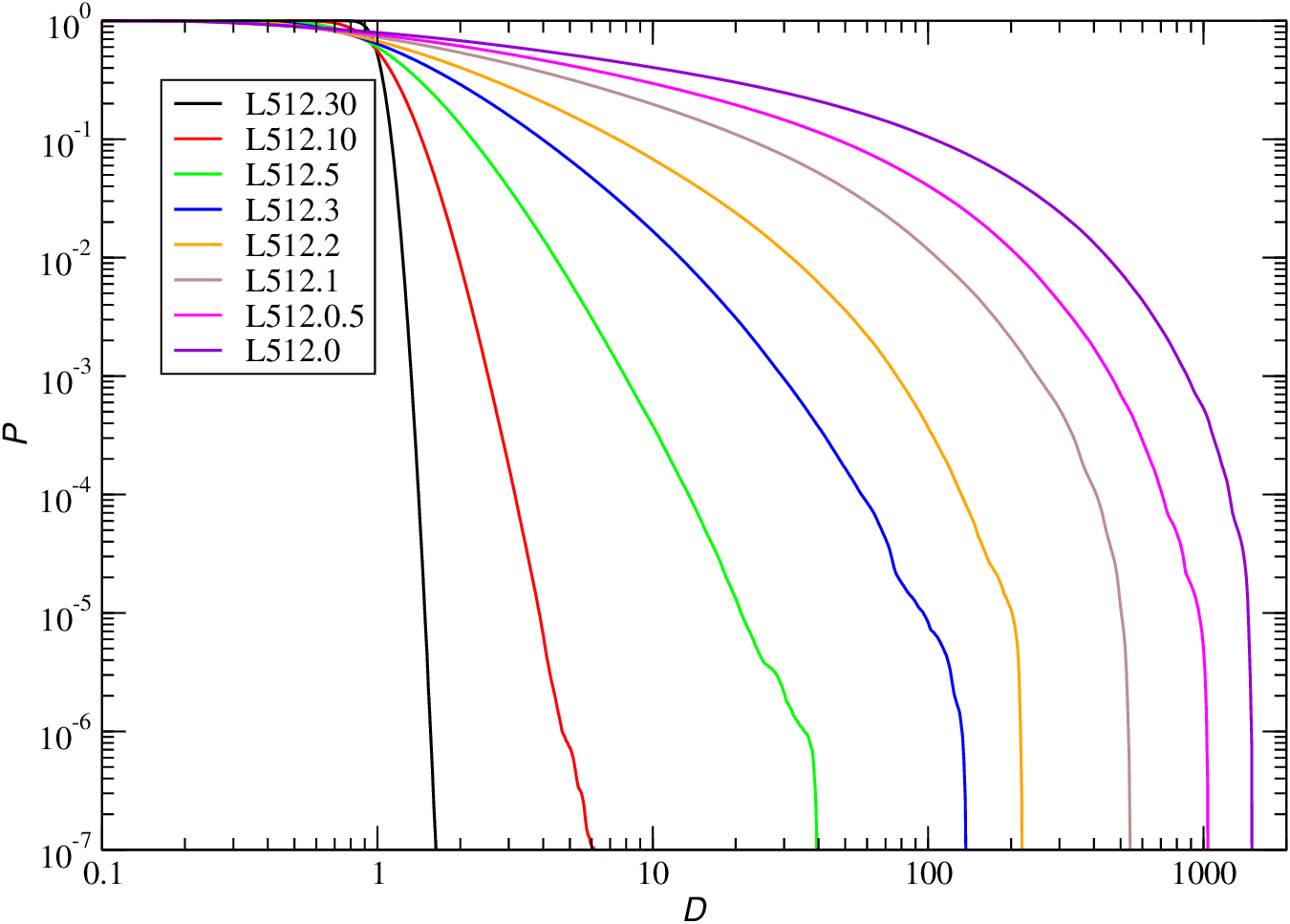}}
\hspace{1mm}
\resizebox{0.32\textwidth}{!}{\includegraphics*{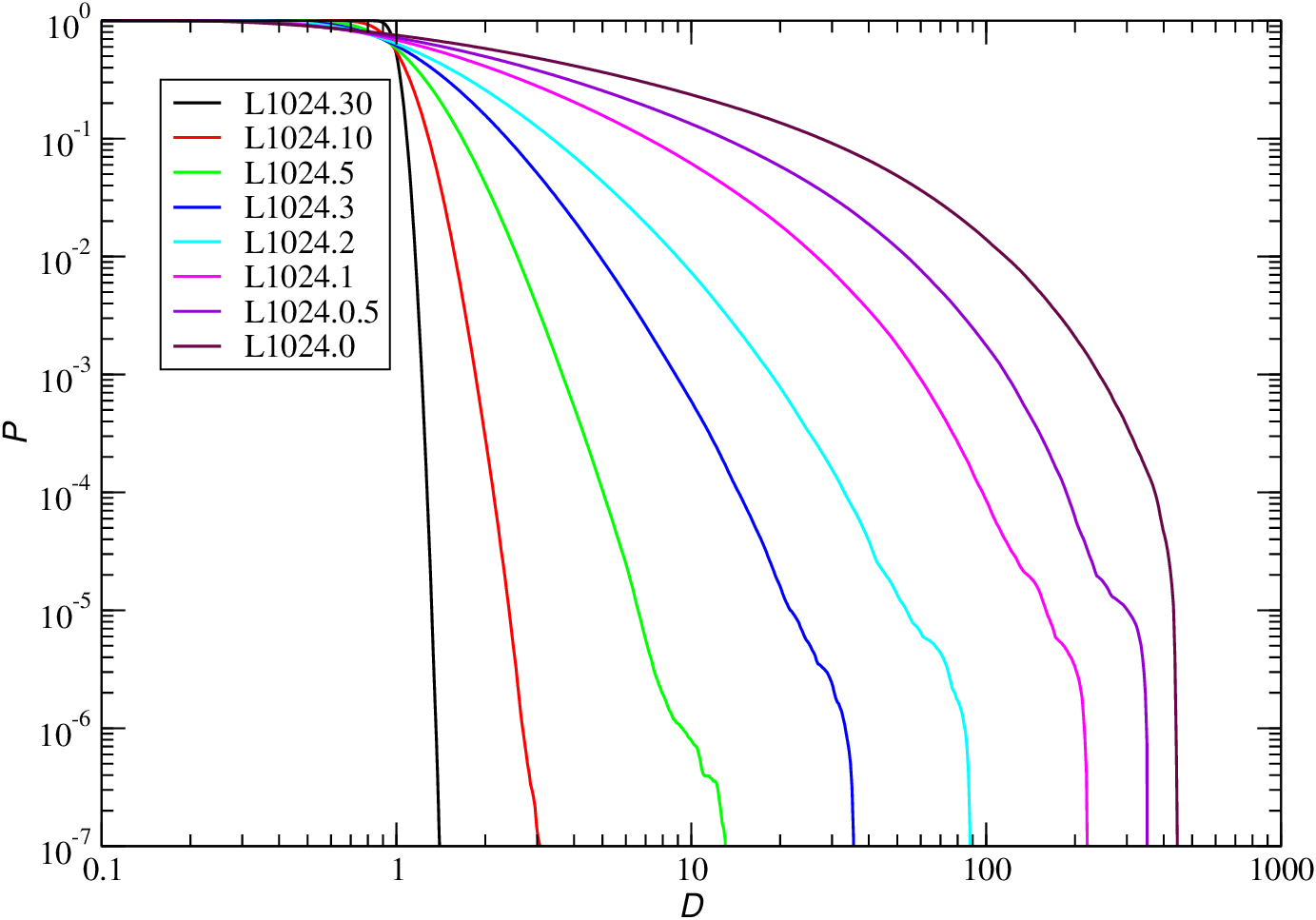}}\\
\hspace{1mm}
\resizebox{0.32\textwidth}{!}{\includegraphics*{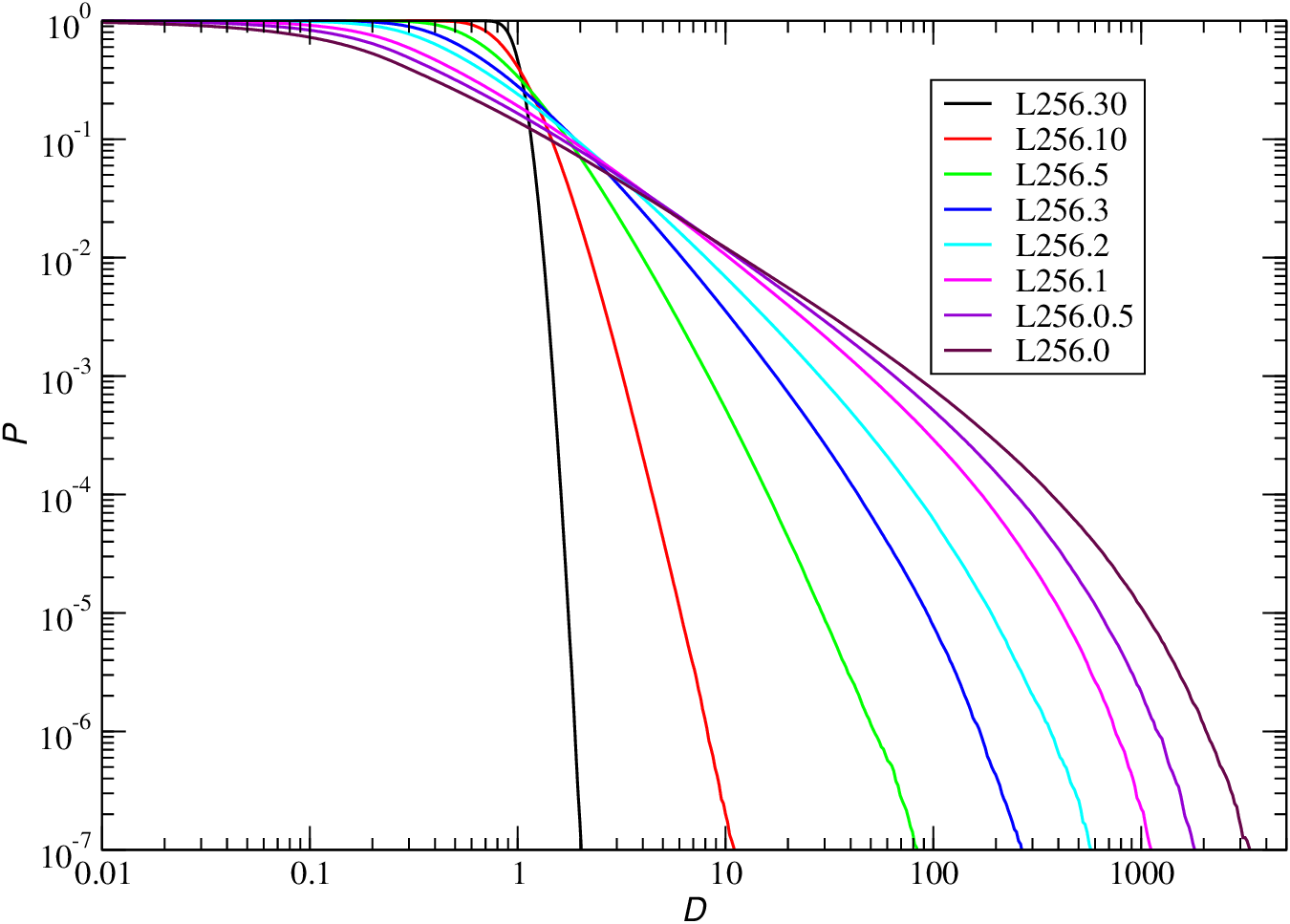}}
\hspace{1mm}
\resizebox{0.32\textwidth}{!}{\includegraphics*{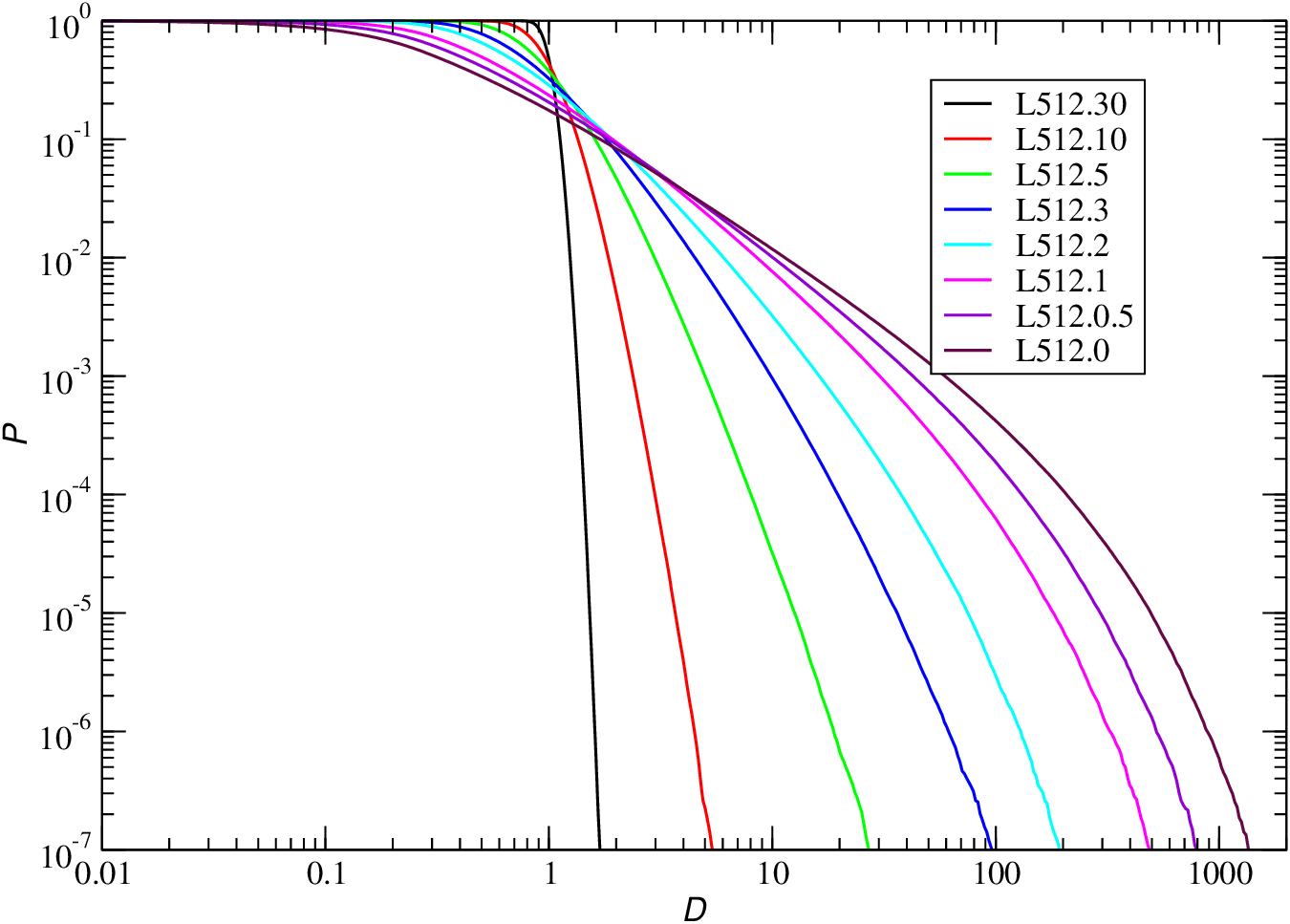}}
\hspace{1mm}
\resizebox{0.32\textwidth}{!}{\includegraphics*{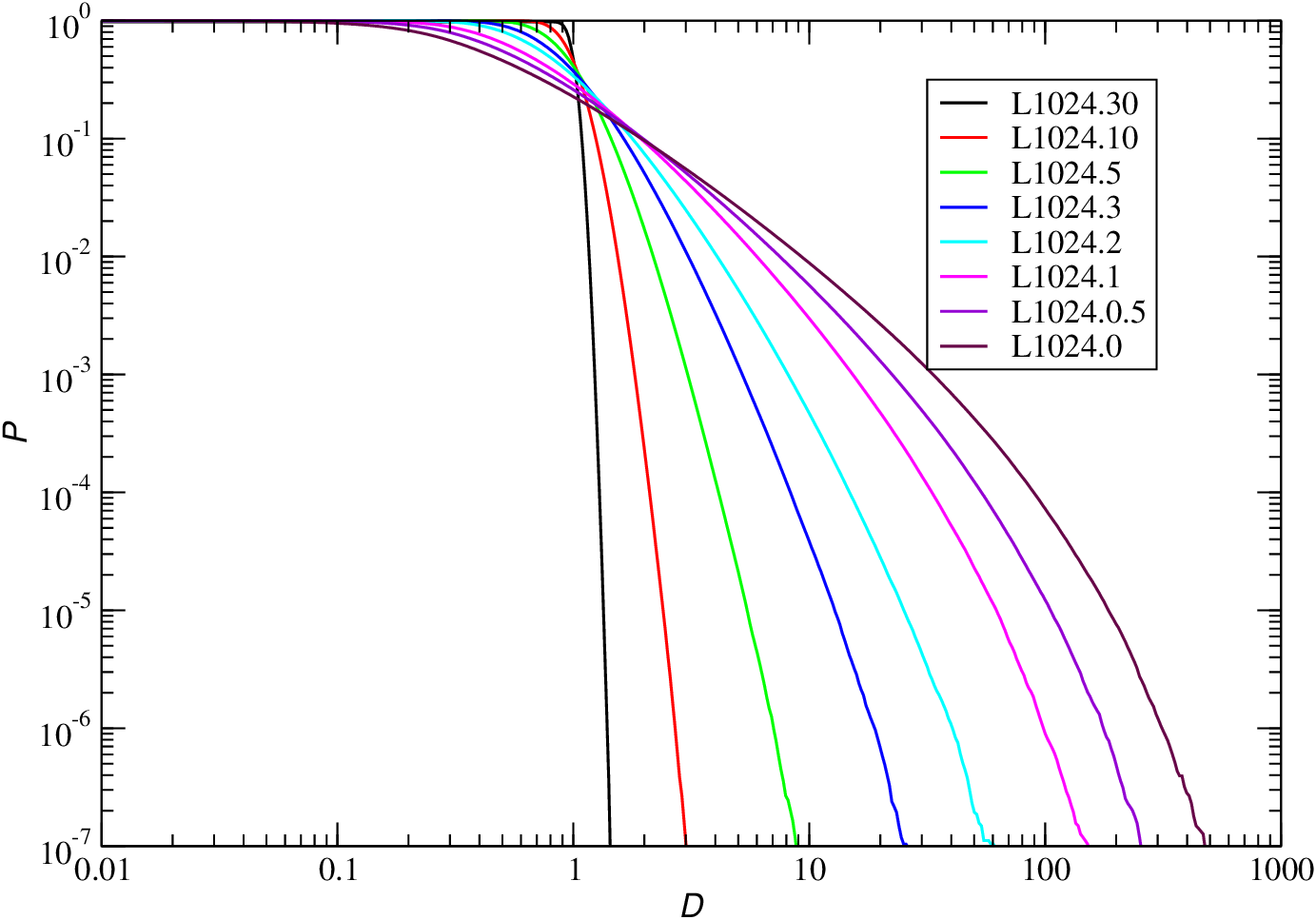}}\\
\hspace{1mm}
\resizebox{0.32\textwidth}{!}{\includegraphics*{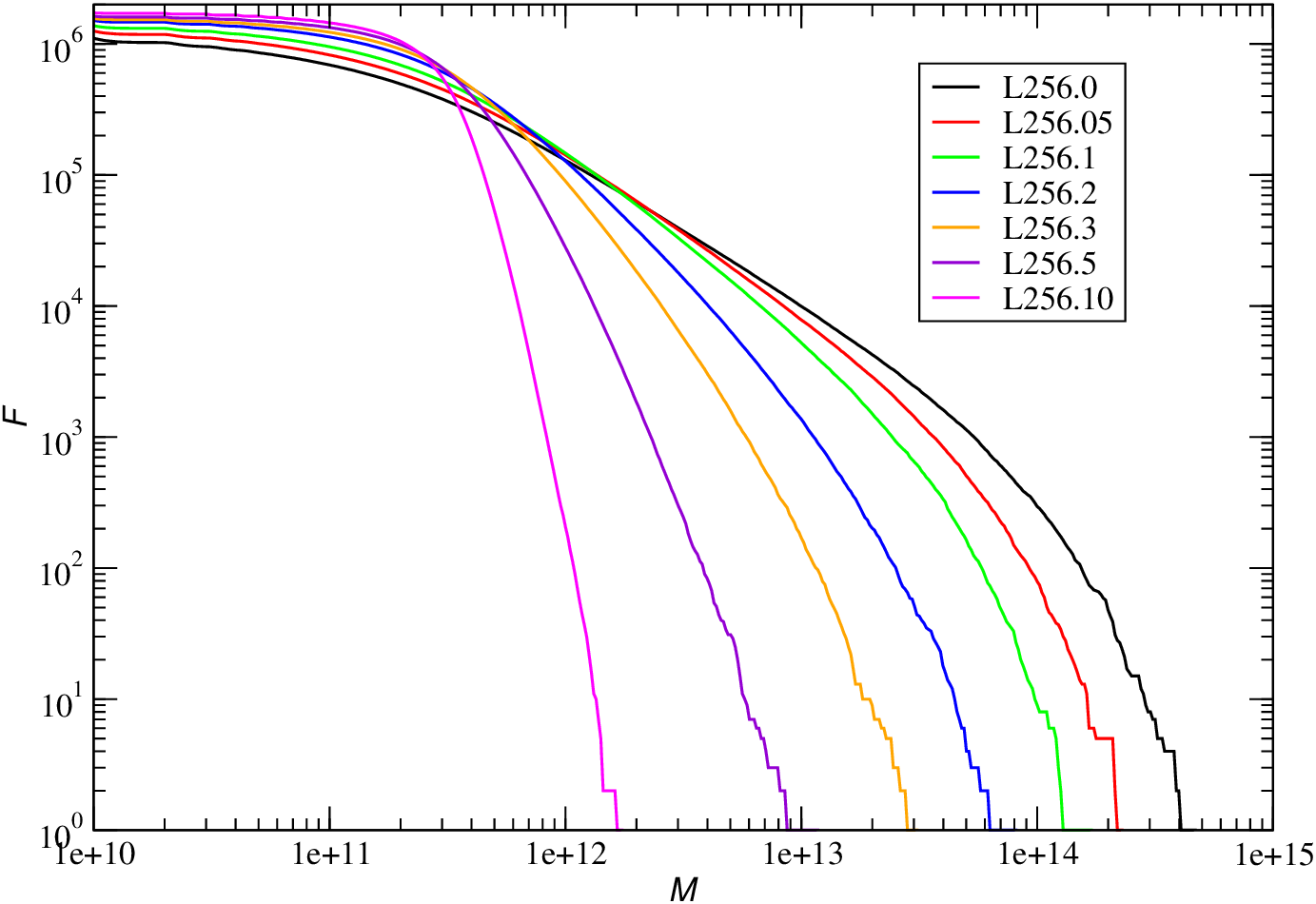}}
\hspace{1mm}
\resizebox{0.32\textwidth}{!}{\includegraphics*{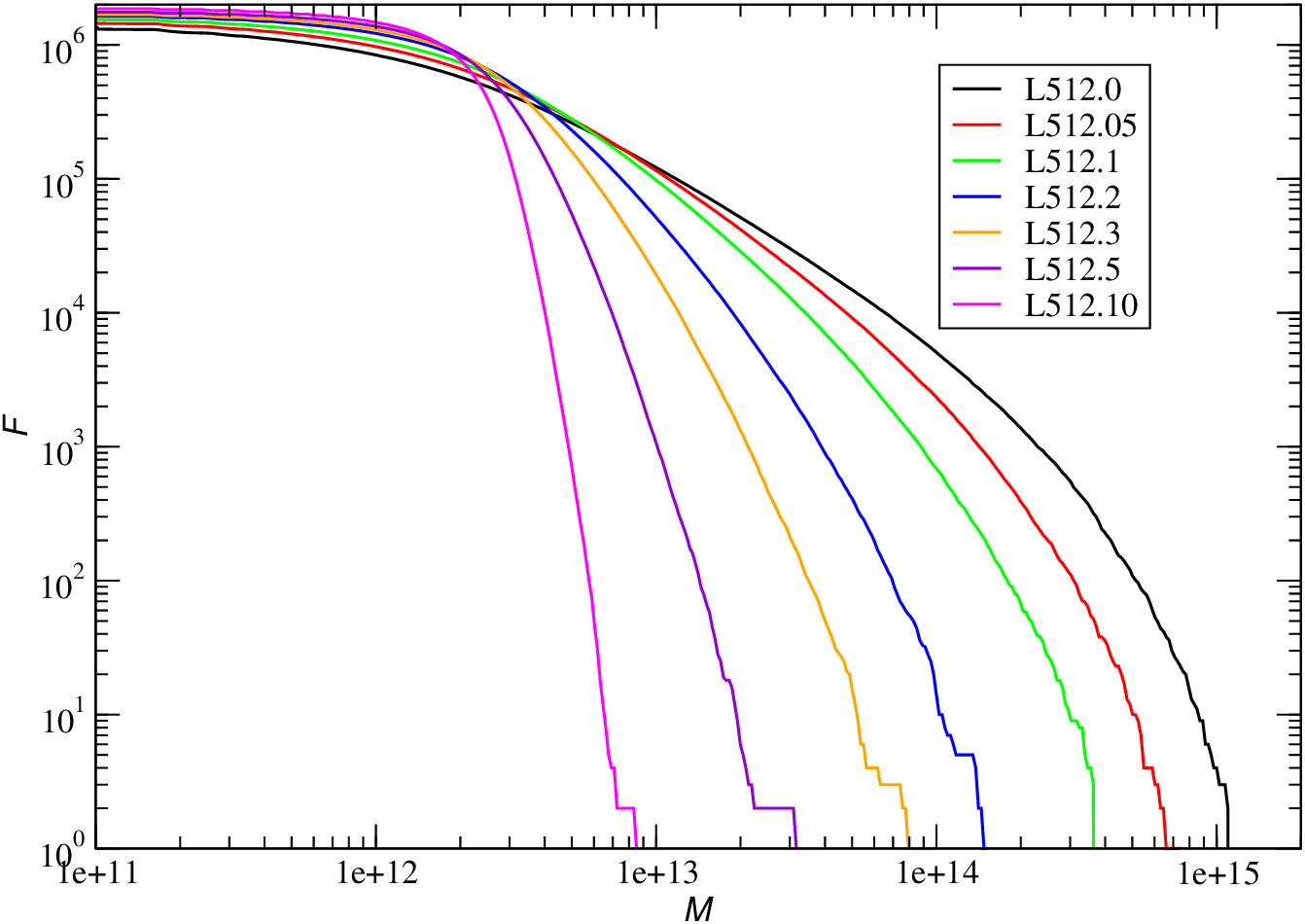}}
\hspace{1mm}
\resizebox{0.32\textwidth}{!}{\includegraphics*{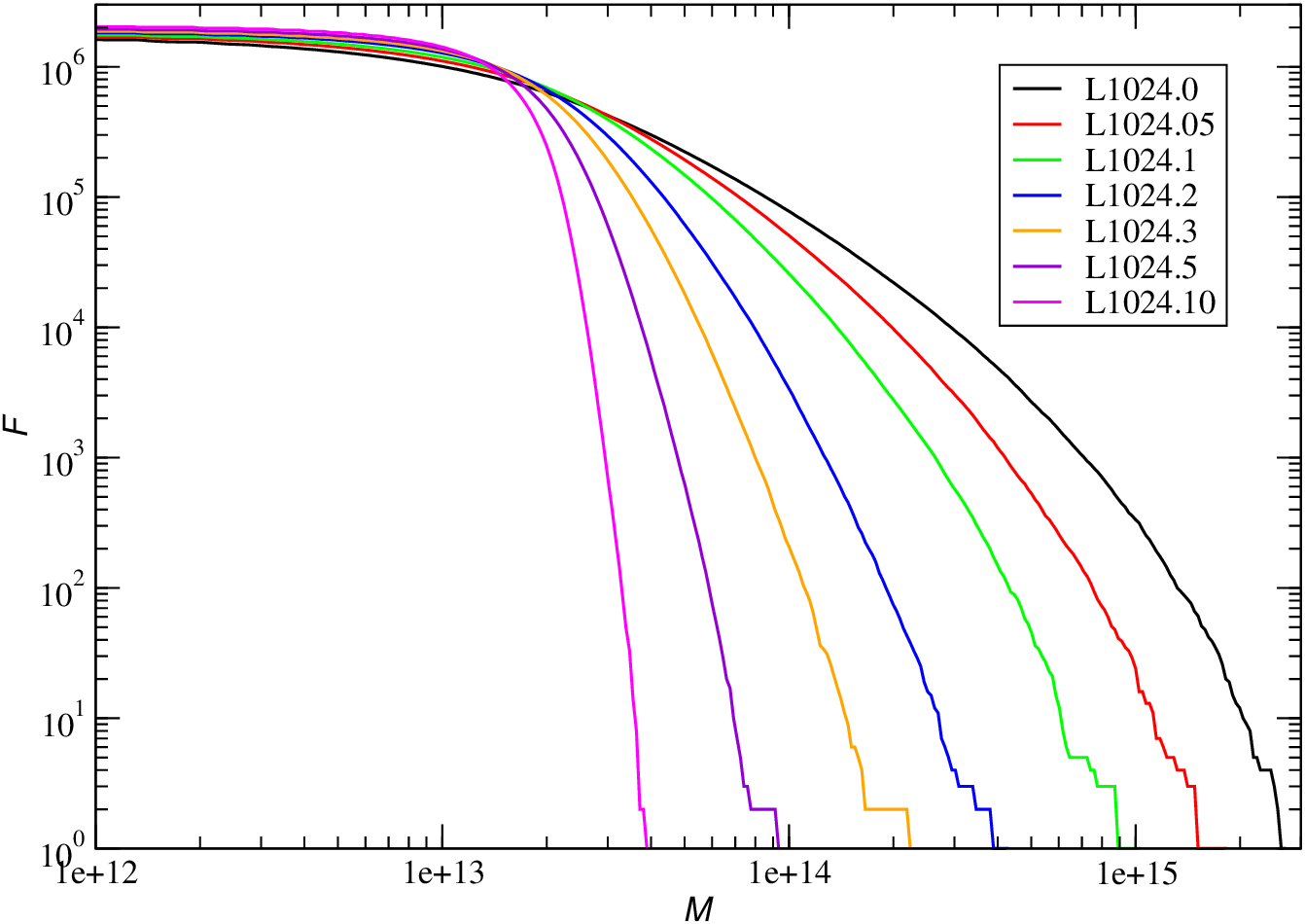}}
\caption{Upper panels: Evolution of cumulative local densities of
  particles. Middle panels: Evolution of cumulative densities of
  density fields of models. Lower panels: Evolution of cumulative
  numbers of model halos. Left, central and right panels are for
  models L256, L512 and L1024, respectively.  In all panels 
   the  second index in  the model name is redshift.  }
\label{fig:Fig2} 
\end{figure*} 

Traditionally the bias parameter $b$ is defined  using
density contrasts:
\be
b = \delta_g/\delta_m,
\label{bias0}
\ee
where $\delta_g$ and $\delta_m$ are density contrasts of galaxies and
matter.  This definition ignores the fact, that in voids the galaxy
density is equal to zero, and the galaxy density contrast
$\delta_g=-1$, which leads to unphysical results. To avoid this
difficulty we define the bias through correlation functions of
galaxies and matter, following the original definition by
\citet{Kaiser:1984}. Also we consider the bias not as a constant, but
as a function of the separation $r$ of galaxies in the correlation
function, $b(r)$.

We calculated the correlation functions $\xi(r)$ for all samples of
$\Lambda$CDM models with a series of particle density limits.  The ratio of
correlations functions of samples with particle density limits
$\rho_0 > 0$ to limit $\rho_0 =0$ (which contains all particles)
defines the bias function $b(r,\rho_0)$:
  \begin{equation}
  b^2(r, \rho_0) = \xi(r,\rho_0)/\xi(r,0).
\label{bias}  
\end{equation}
Bias depends on the luminosity of galaxies, in our case on the
particle density threshold $\rho_0$, used in the calculation of
correlation functions. 

Bias functions have a plateau at $6 \le r \le 20~\Mpc$, see
Fig.~\ref{fig:Fig5} below.  This feature is similar to the plateau around
$k \approx 0.03$~$h$~Mpc$^{-1}$ of relative power spectra
\citep{Einasto:2019aa}.  Following  \citet{Einasto:2020aa,
  Einasto:2021ti} we use this plateau to measure the relative amplitude 
of the correlation function, i.e. of the bias function, as the bias
parameter,
\be
b(\rho_0)= b(r_0,\rho_o),
\label{bias2}
\ee
where $r_0$ is the value of the separation $r$ to measure the
amplitude of the bias function.  We calculated for all samples bias
parameters for two values of the comoving separation:
$r_0=r_{6}=6~\Mpc$, and $r_0=r_{10}=10~\Mpc$, as functions of the
particle density limit, $\rho_0$.  The comoving separation
$r_{6}=6~\Mpc$ was used by \citet{Einasto:2021ti} to estimate bias
parameters for the present epoch $z=0$. The present analyse suggested
that for earlier epochs the value $r_{10}=10~\Mpc$ is preferable.  At
smaller distances, bias functions are influenced by the distribution
of particles in halos, and at larger distances, the bias functions
have wiggles, which makes difficult the comparison of samples with
various particle density limits.

\section{Evolution of particle densities and of the density field}

In this Section we compare first the evolution of the variance of density
perturbations.  Thereafter we discuss the evolution   of particle
densities and density fields with the cosmic epoch $z$, and density
distributions of biased model samples. Finally we describe the
evolution of density field halos.

\subsection{Evolution of the variance of density perturbations with
  cosmic epoch $z$}

We found density fields of models L256, L512 and L1024 for all
epochs for which we have computer outputs.  Using these fields we
calculated the second moment  or variance $\sigma^2$ of density fields. It
is useful to call $\sigma$ as the dispersion of the 
density field.   The evolution of $\sigma$ for various smoothing
parameter $R$ values  is shown in left panel of 
Fig.~\ref{fig:Fig1}.  We see that the growth of the density 
dispersion with cosmic epoch is approximately  linear in $\log(\sigma)$ --
$\log(1+z)$ diagram for smoothing length $R=8~\Mpc$.   For smaller
smoothing lengths deviations from a linear growth in smaller redshift
$z$ region are visible.

It is remarkable that the growth of $\sigma$ of different models but
with identical smoothing length $R$ are very close.  The dispersion
$\sigma_8$ found with $R=8~\Mpc$ using the $B_3$-spline smoothing is
given in Table~\ref{Tab1}. For all models is is close to value
$\sigma_8=0.64$.  All models were calculated with initial dispersion
parameter $\sigma_8=0.84$, which corresponds to the linear evolution
of density perturbations. As we see, the actual value of $\sigma_8$ is
slightly lower than the linearly extrapolated value. What is important
for the use of our models is the fact, that actual density dispersions
of all models for identical smoothing lengths are very close to each
other.

Right panel of Fig.~\ref{fig:Fig1} presents the dependence of the
growth of the dispersion of density fluctuations on smoothing length
$R$ and redshift $z$.  This dependence is also almost linear in logarithmic scale.

\begin{figure*}
\centering 
\hspace{1mm}  
\resizebox{0.35\textwidth}{!}{\includegraphics*{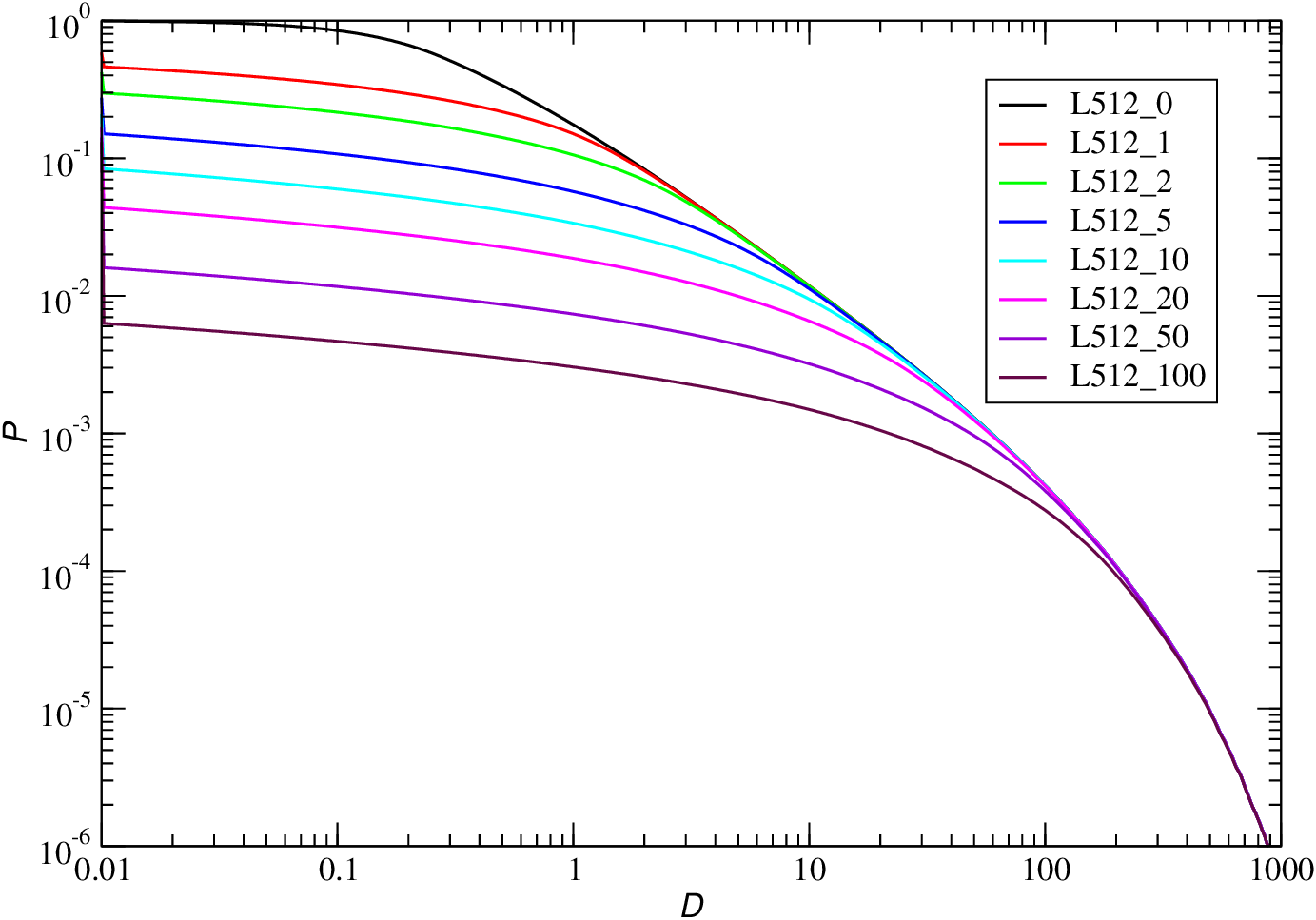}}
\hspace{1mm}  
\resizebox{0.35\textwidth}{!}{\includegraphics*{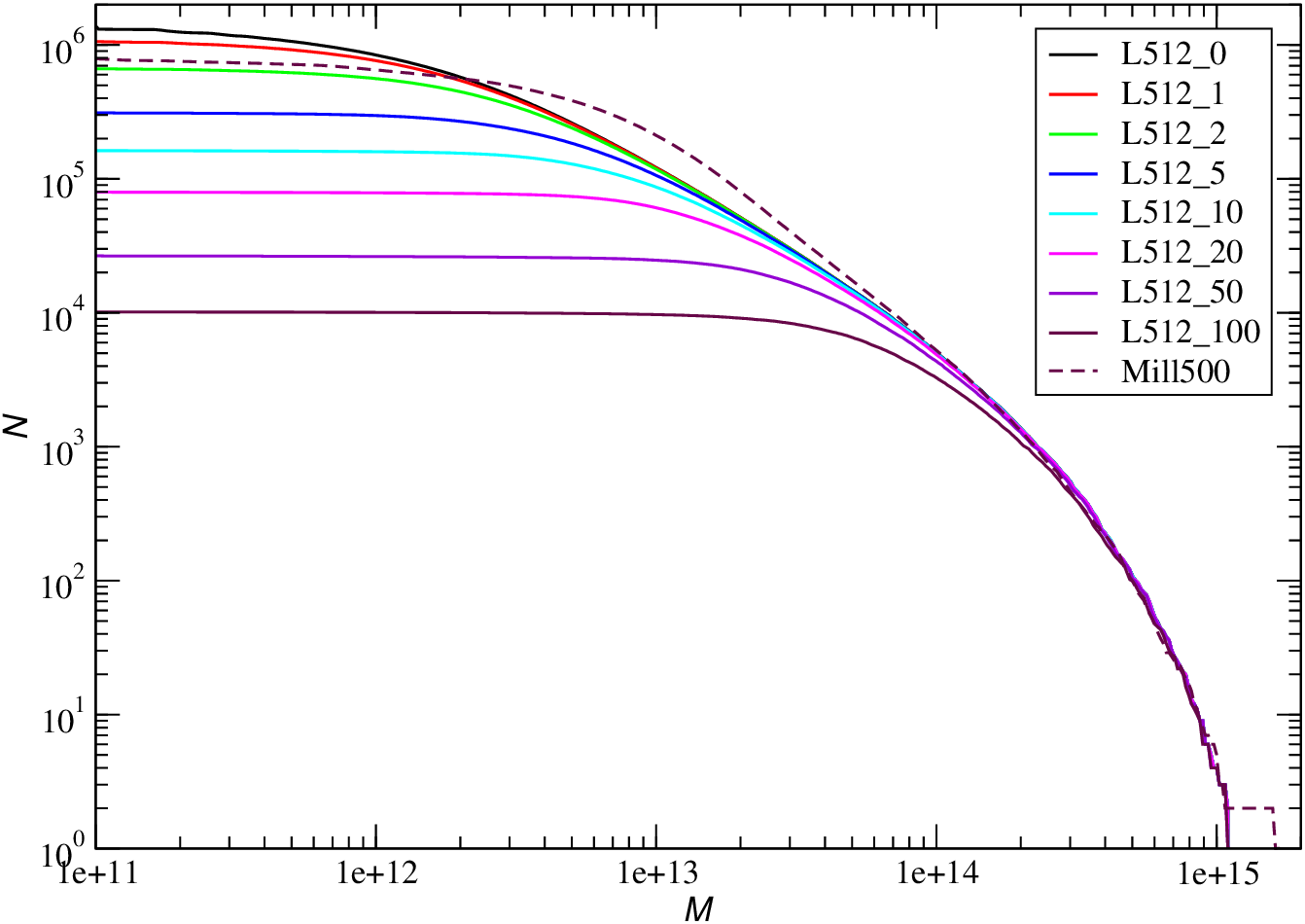}}
\caption{Left: Cumulative density distributions of biased L512 model
  samples, normalised to the total number of particles.  Right:
  Cumulative numbers of density field halos of biased L512 model
  samples.  Both distributions are given for various particle density
  limits $\rho_0$, indicated as symbol label.  For comparison we show
  by dashed line in the right panel the cumulative distribution of halo
  mass of the Millennium simulation. All distributions are for the
  present epoch $z=0$.  }
\label{fig:Fig3} 
\end{figure*}

\subsection{Evolution of particle densities and density fields with cosmic epoch $z$}

During the simulations of $\Lambda$CDM models we found for each particle the
local density.  This local density depends on 
the resolution of simulation, and has an effective smoothing length
$R_0=L_0/N_{\mathrm{grid}}$ in units $\Mpc$, given in Table ~\ref{Tab1}.  This local density
$\rho_0$ was used to select particles to form biased model samples,
which correspond to simulated galaxy samples.  We show in upper panels
of Fig.~\ref{fig:Fig2} the evolution of cumulative local densities of
particles with cosmic epoch $z$ for our models L256, L512 and L1024,
normalised to the total number of particles.  This Figure allows to
see the range of possible density limits $\rho_0$ to select particles
for simulated galaxy samples at particular epoch.  

For comparison we show in middle panels of Fig.~\ref{fig:Fig2} the
evolution of density fields with cosmic epoch of the same models.  The
comparison of upper and middle panels of Fig.~\ref{fig:Fig2} shows
that distributions of particle densities are different from the
distribution of densities of respective density fields -- cumulative
densities of particles are much higher than cumulative densities of
density fields. This difference increases with cosmic epoch
(decreasing $z$ value).  The reason for these differences lies in the
volume occupied by particles in high-density regions, which is much
smaller than the fraction of these particles in the total sample of
all particles.

Fig.~\ref{fig:Fig2} shows also that distributions of particle
densities and density fields depend on the model size  -- upper
limits of particle densities and density field values of the model
L256 are higher than those of models L512 and L1024.  This difference
is due to various resolution of models -- the effective smoothing
length of the model L256 is $0.5~\Mpc$, of the model L512 it is
$1~\Mpc$, and of the model L1024 it is $2~\Mpc$, as shown in
Table~\ref{Tab1}. 

Fig. ~\ref{fig:Fig2} shows that for the simulation L256 we can use  particle
density limits for biased samples, $\rho_0 \le 100$, only for epochs
$z\le 5$, since at epoch $z \ge 5$ there are no particles with $\rho
\ge 100$.  Thus for this simulation and epoch $z \ge 10$ we can use only
particles with density limit $\rho_0 \le 10$.  The range of possible
density limits for simulations L512 and L1024 is smaller.

We show in the left panel of Fig.~\ref{fig:Fig3} cumulative densities
of particles for biased samples of the simulation L512 for various
particle density limit $\rho_0$ at the present epoch $z=0$.  The upper
curve presents the cumulative densities of particles of the whole
unbiased model L512.00, following curves show distributions for biased
models with particle density limits up to $\rho_0=100$. We see that at
highest densities all curves coincide, and that curves for biased
samples deviate from the unbiased model at densities, approximately
equal to the limit $\rho_0$.

\subsection{Evolution of halos of  density fields}

In this paper we used particle density limited samples to simulate
galaxies. An alternative is to use density field halos.  We
calculated for all models and simulation epochs density field halos.
Usually halos are formed using simulation particles. Another 
possibility is to use high-resolution density fields.  We developed a
simple halo finding algorithm, which finds peaks of density
fields. Halos are formed by adding densities of surrounding 27 cells,
including the central peak cell. This halo finding algorithm is very
fast. The cumulative distribution of halos masses of all three models
and simulation epochs is shown in bottom panels of
Fig.~\ref{fig:Fig2}.  Cumulative halo mass functions of biased L512
models for the present epoch are shown in Fig.~\ref{fig:Fig3}.

The comparison of upper and lower panels of Fig.~\ref{fig:Fig2} shows
that cumulative distributions of local densities of particles and
halos of various masses have some similarity for all simulation
epochs.  However, there exist differences, which are largest at low
density and halo masse ranges.  These difference are due to the fact
that our halos do not contain subhalos -- halos sum densities within
the whole region $\pm 1$ cells around the central peak.

Density field halos can be found also using catalogs of simulated
galaxies.  To test this possibility we calculated density fields for
Millennium simulation \citep{Springel:2005aa} galaxies by
\citet{Croton:2006aa}.  Fig.~\ref{fig:Fig3} shows by dashed line the
cumulative distribution of masses of Millennium density field halos.
Halos were found using the same procedure as for $\Lambda$CDM model
halos. Millennium simulation has the box size $500~\Mpc$, very close
to out L512 model, thus results are comparable.  We see that in
high-mass halo region distributions of our L512 model and Millennium
simulation coincide.  In small mass region our L512 model yields 
higher number of halos, since it uses all particles.

We see that the use of halos instead of particles can be applied to study
the evolution of the bias parameter. However, in this case the whole
analysis should be made using halos.  The comparison of halo and
particles distributions  shows, that the use of all particles yields
more detailed   information on the internal structure of   halos. For
this reason we use for the detailed analysis only particle density
selected biased model samples. As shown by
\citet{Einasto:2019aa}, particle density 
limited samples and ordinary halo mass selected samples yield density
fields, very similar to density fields of real luminosity limited SDSS
samples.

\begin{figure*}
\centering 
\hspace{1mm}  
\resizebox{0.32\textwidth}{!}{\includegraphics*{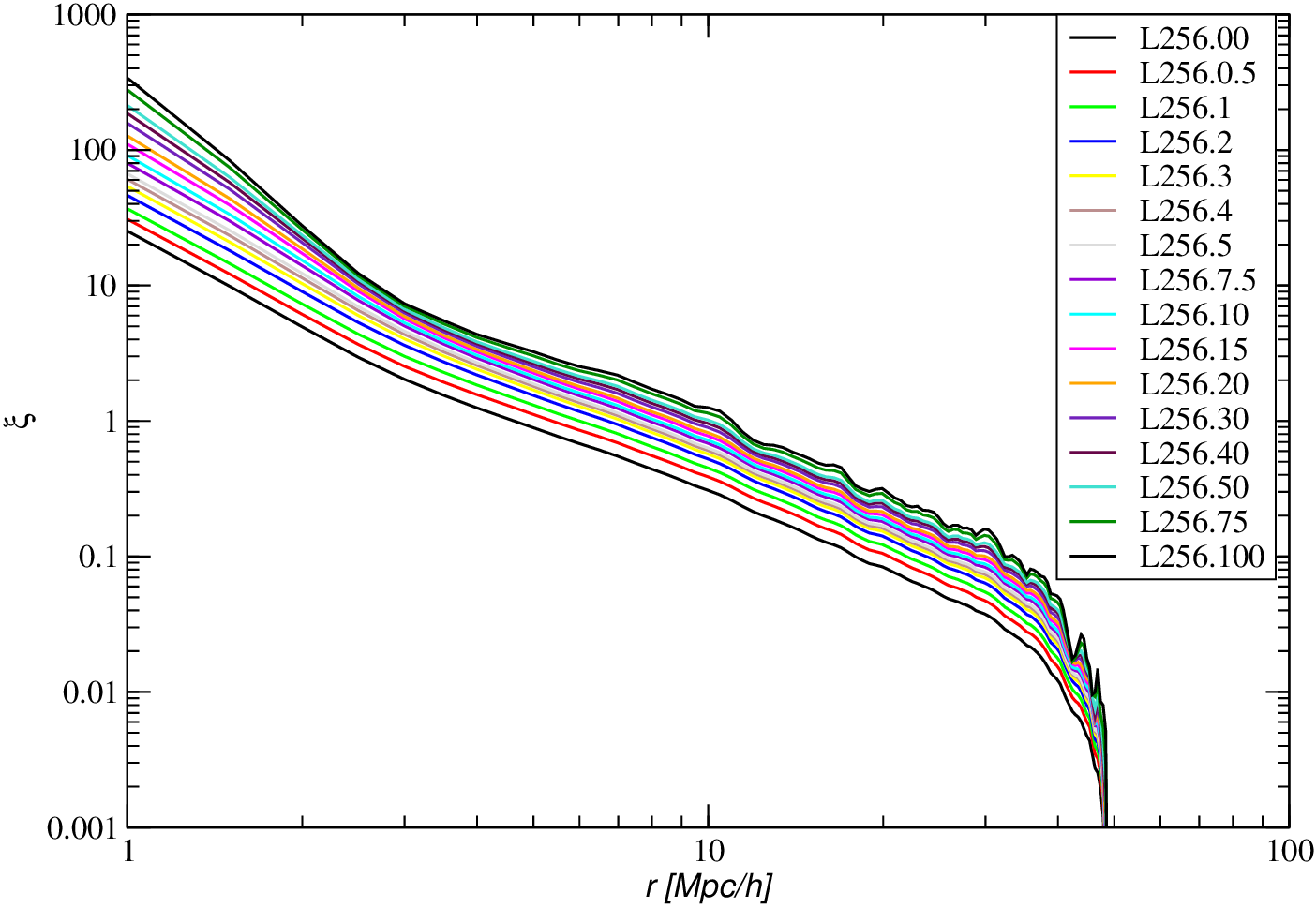}}
\hspace{1mm}  
\resizebox{0.32\textwidth}{!}{\includegraphics*{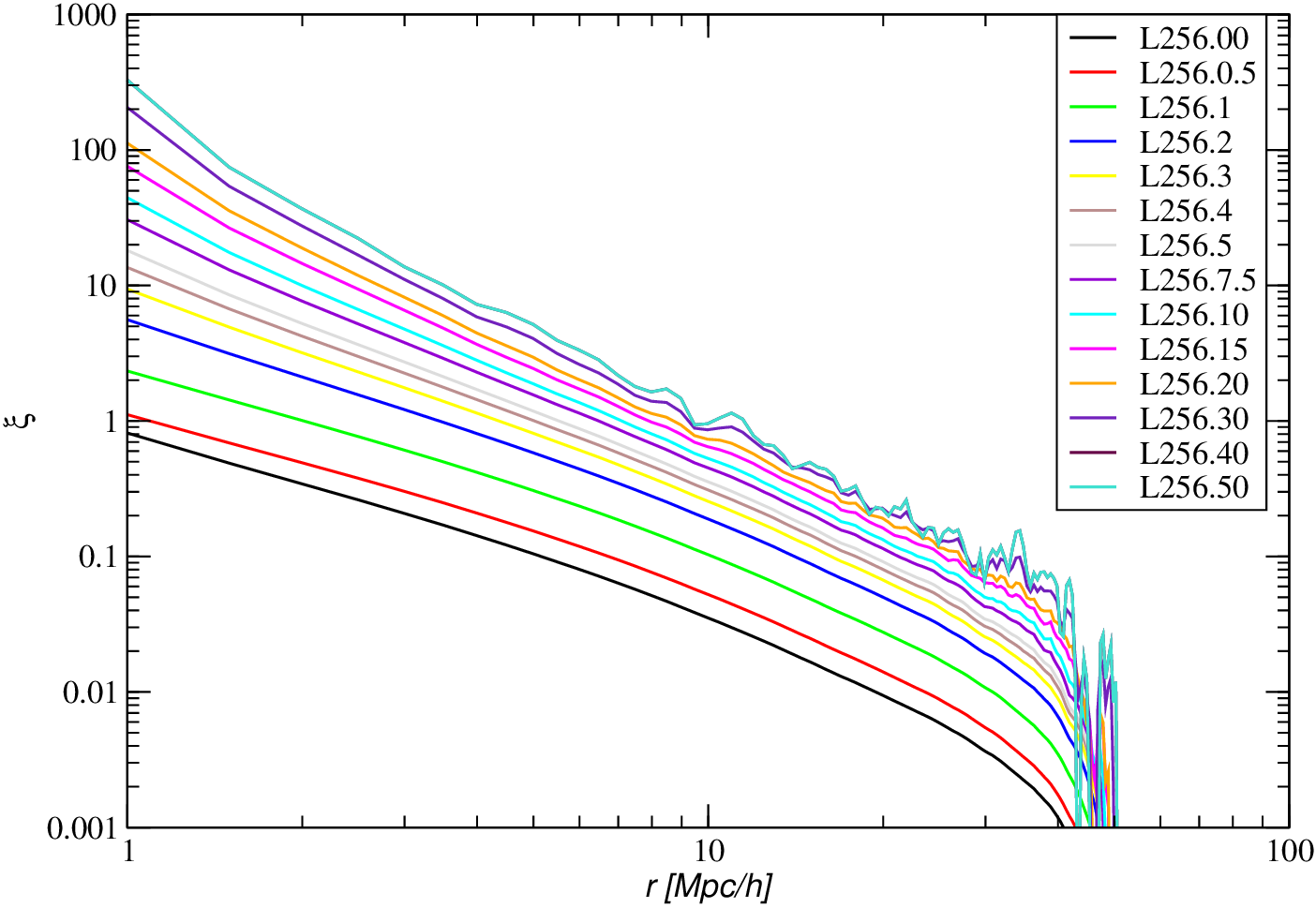}}
\hspace{1mm}  
\resizebox{0.32\textwidth}{!}{\includegraphics*{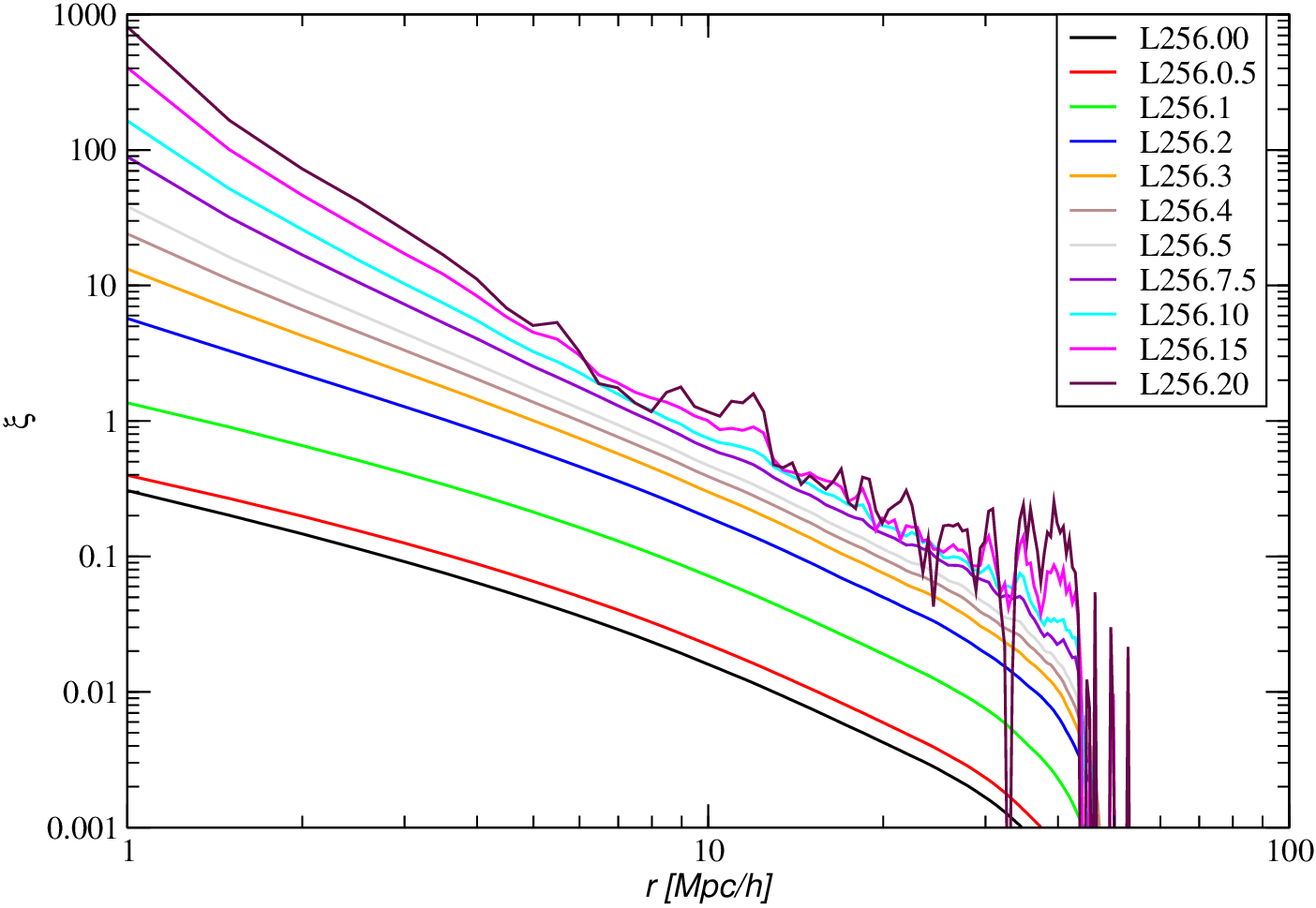}}\\
\hspace{1mm}  
\resizebox{0.32\textwidth}{!}{\includegraphics*{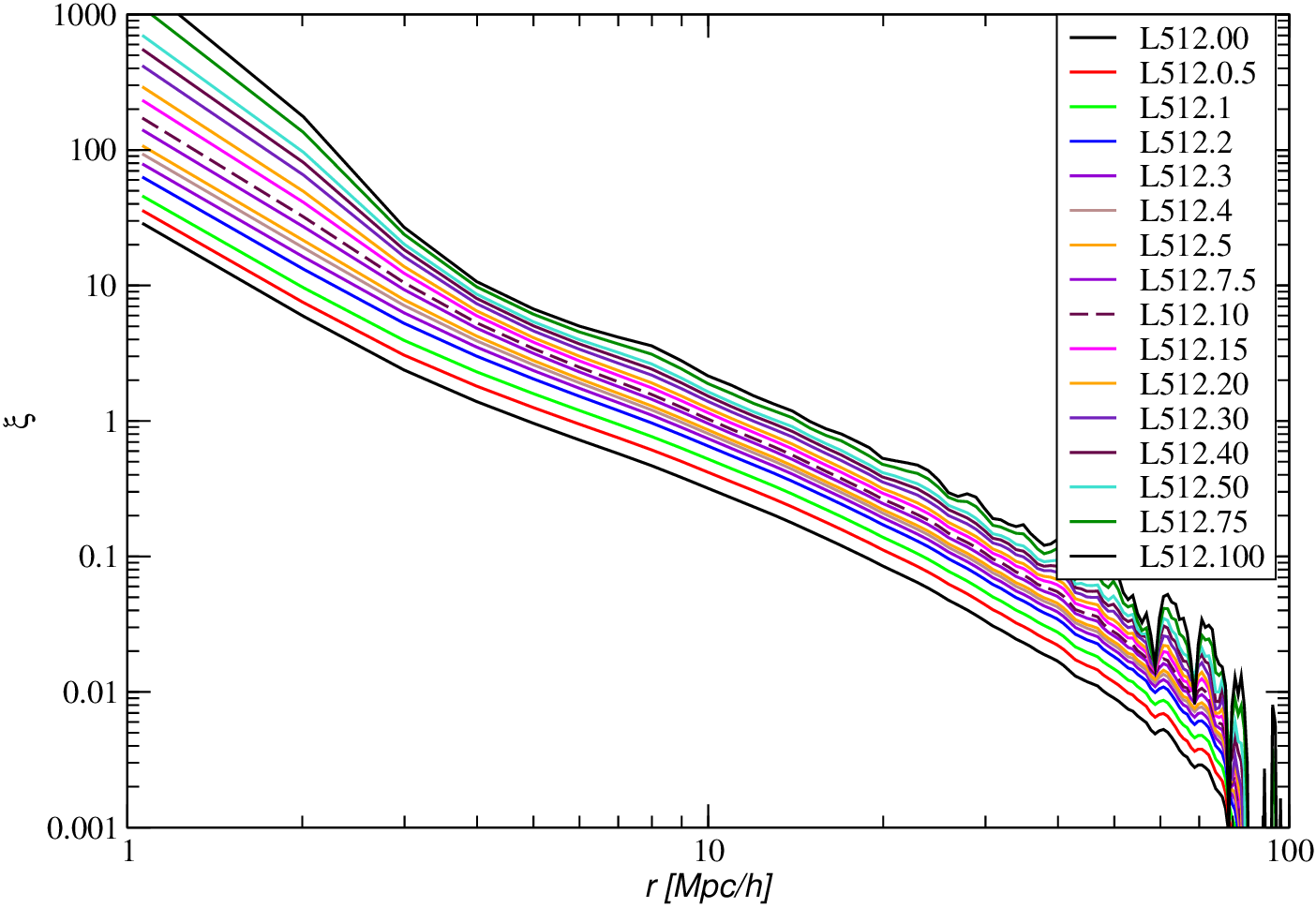}}
\hspace{1mm}  
\resizebox{0.32\textwidth}{!}{\includegraphics*{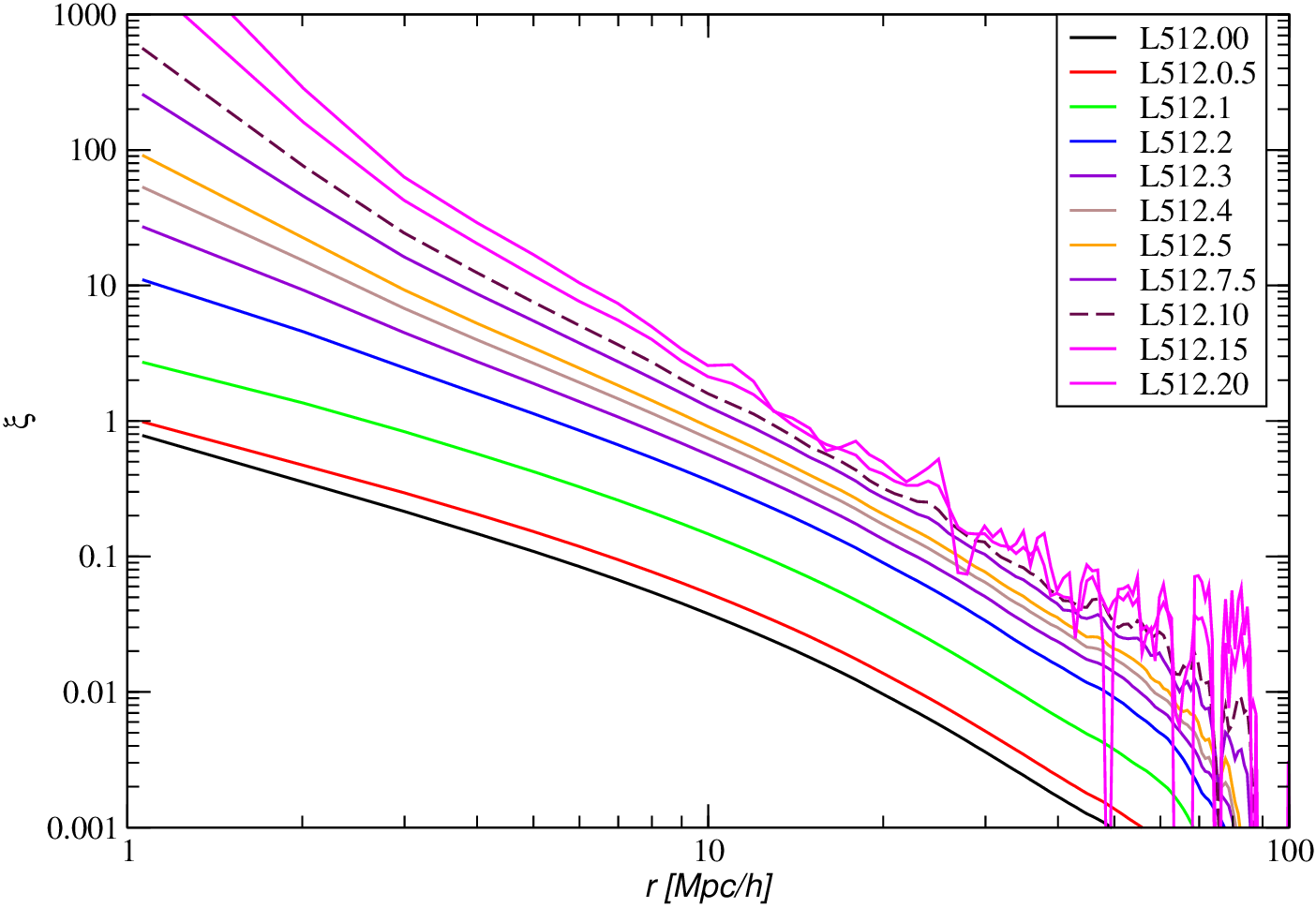}}
\hspace{1mm}  
\resizebox{0.32\textwidth}{!}{\includegraphics*{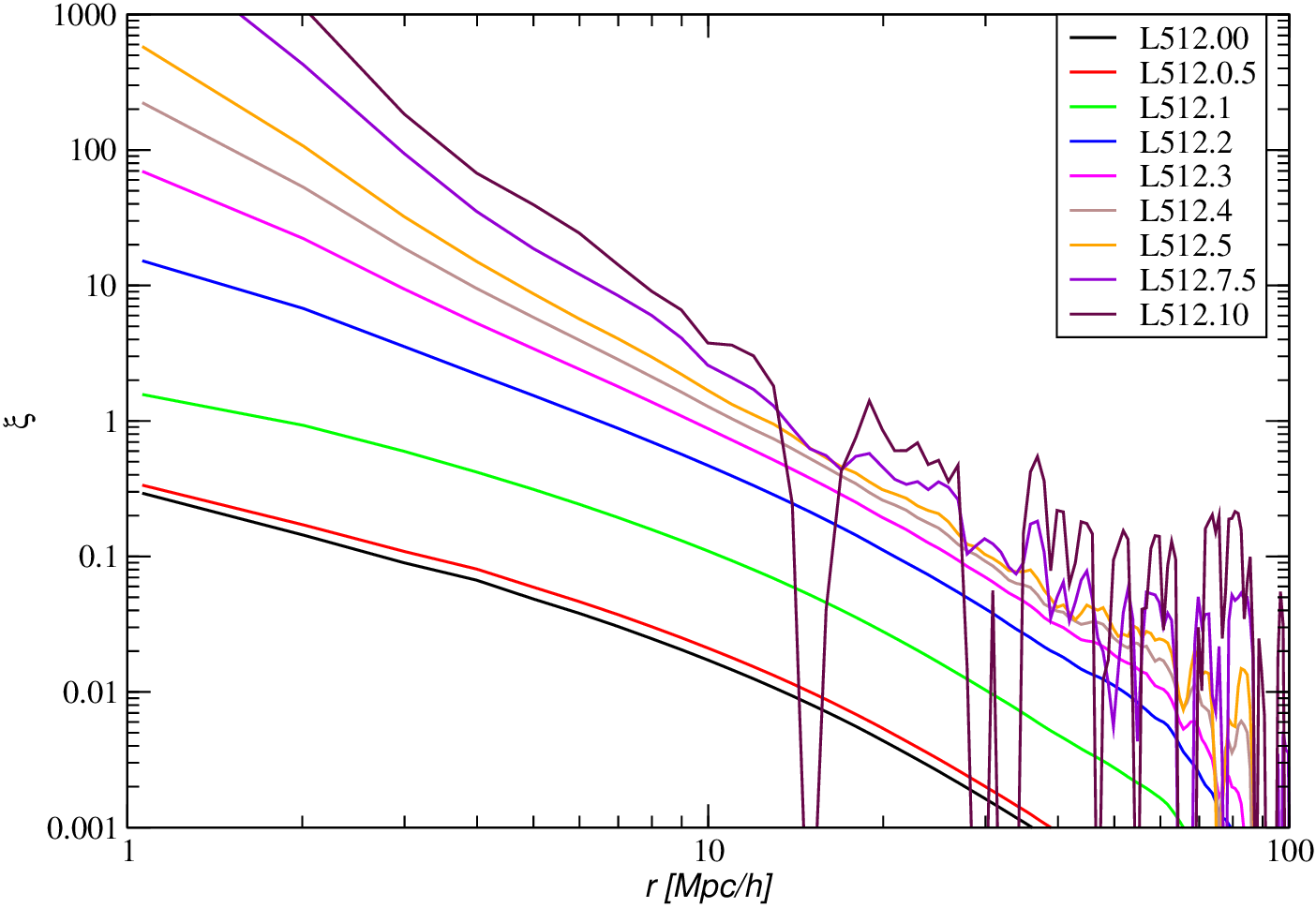}}\\
\hspace{1mm}  
\resizebox{0.32\textwidth}{!}{\includegraphics*{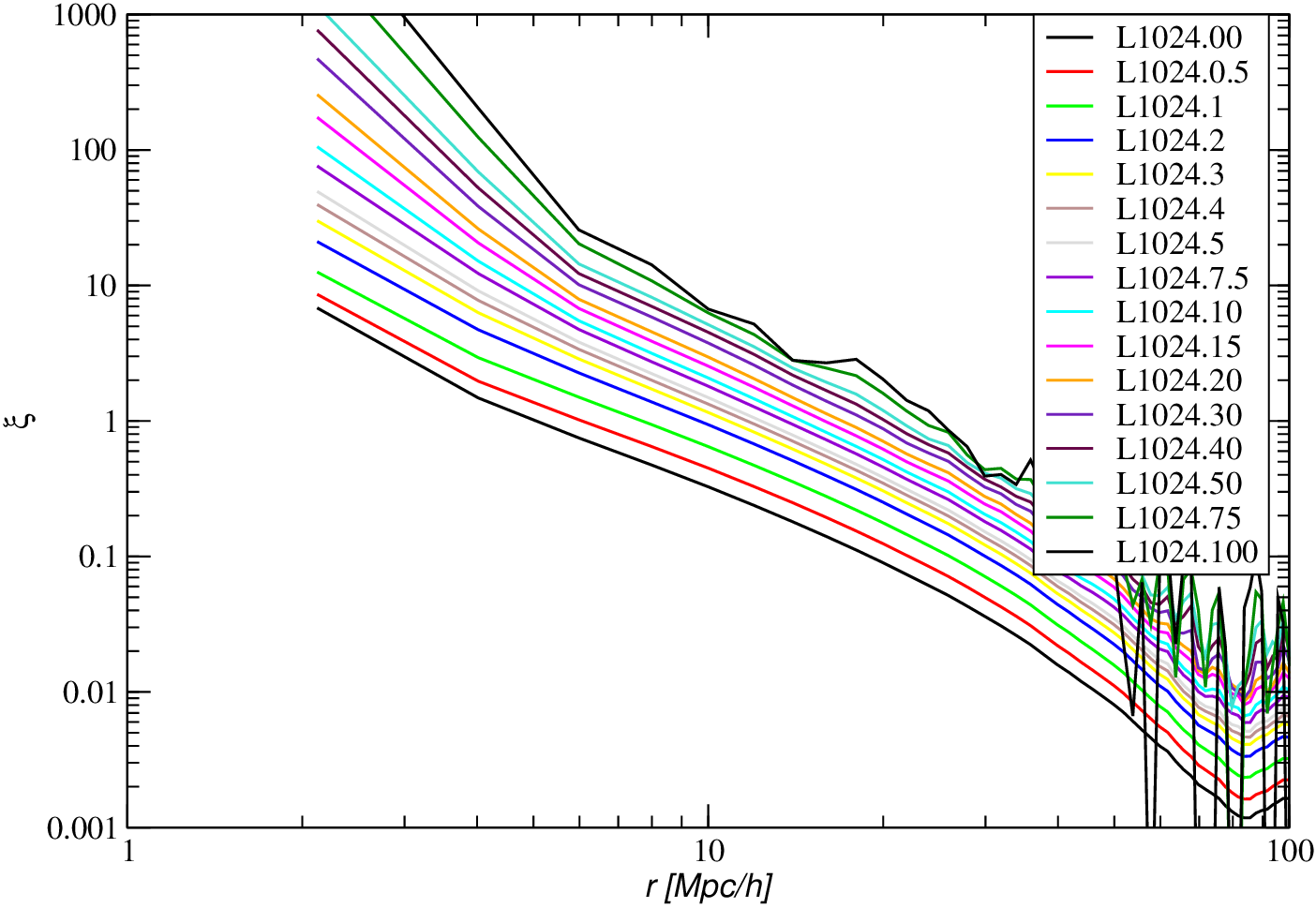}}
\hspace{1mm}  
\resizebox{0.32\textwidth}{!}{\includegraphics*{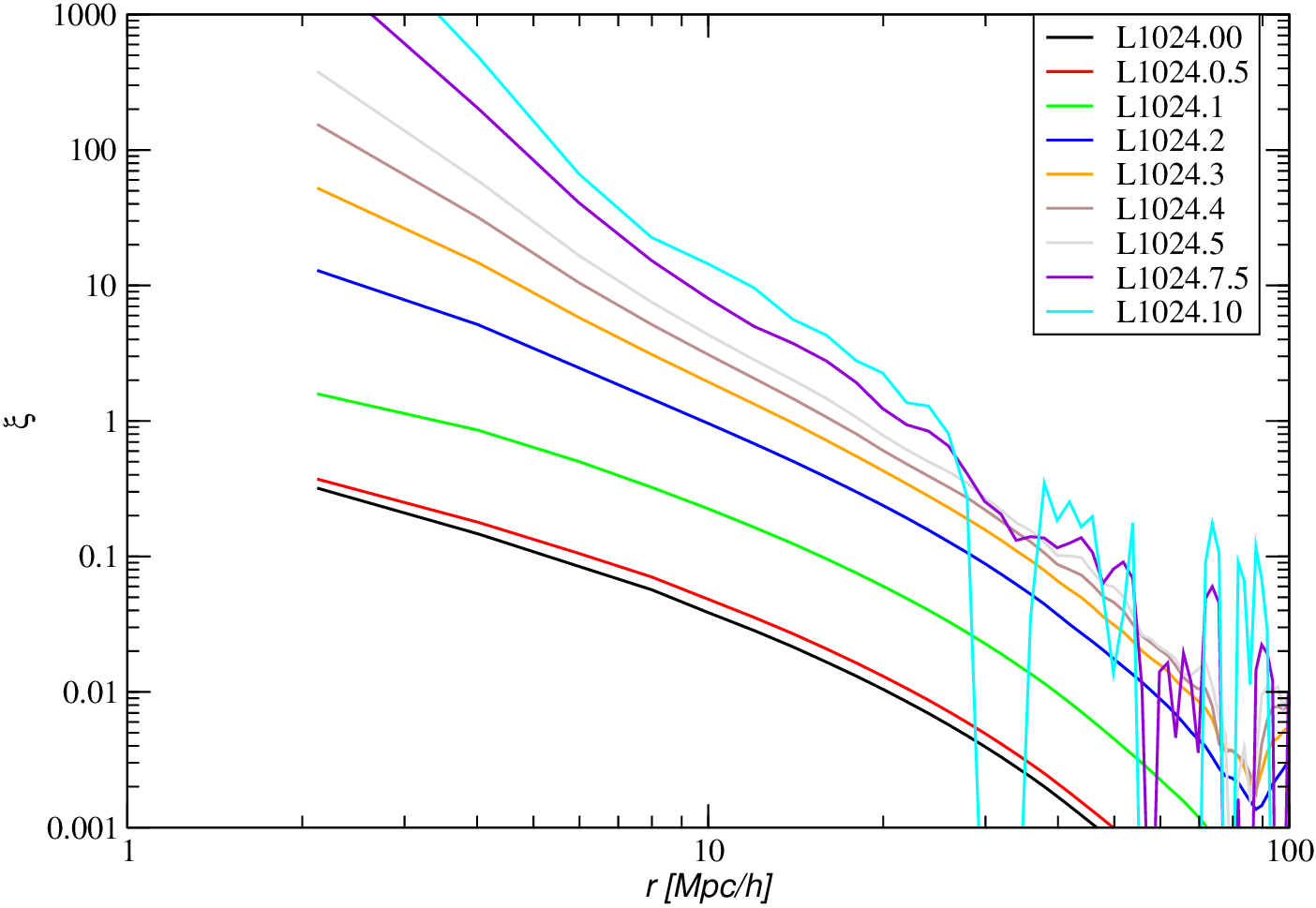}}
\hspace{1mm}  
\resizebox{0.32\textwidth}{!}{\includegraphics*{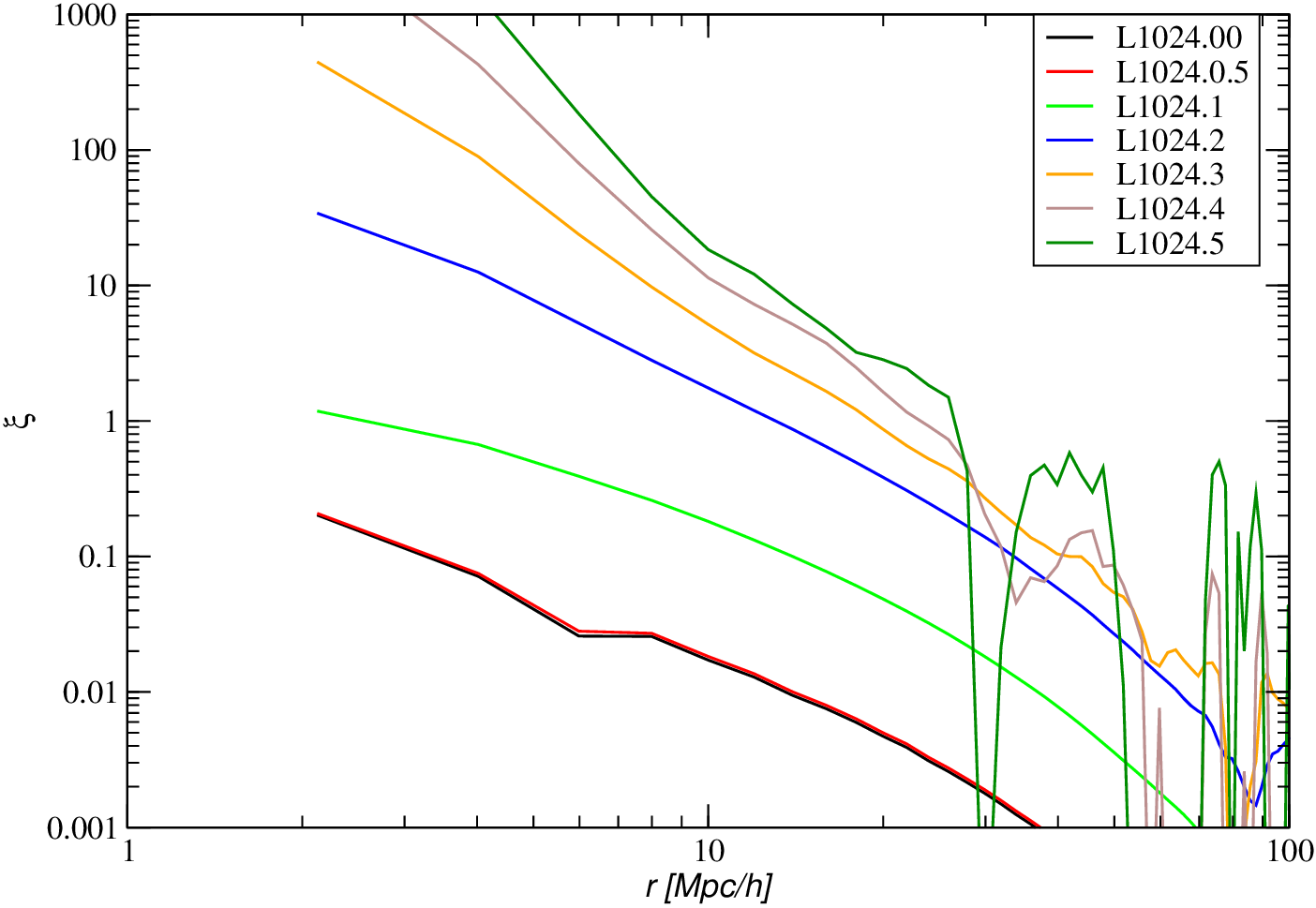}}\\
\caption{Correlation functions of galaxies, $\xi(r)$ for epochs
  $z=0,~3,~5$, shown respectively in the left, central and right
  panels. Upper panels are for the model L256,  middle panels  for the
  model L512, and lower panels  for the L1024 model.  Density limits
  $\rho_0$ are indicated as symbol labels.  }
\label{fig:Fig4} 
\end{figure*}

\section{Evolution of biasing properties of $\Lambda$CDM models}

In this Section we consider the evolution of correlation and bias
functions of biased (simulated galaxy) $\Lambda$CDM models. Next we
describe the evolution of bias parameters with cosmic epoch.  Bias
parameters were derived for various particle density limits. In order
to get bias parameters for comparable objects we calculated bias
parameters for identical fractions of galaxies in samples. The Section
ends with the error analysis.

\subsection{Evolution of  correlation functions of $\Lambda$CDM models}

\begin{figure*}
\centering 
\hspace{1mm}  
\resizebox{0.32\textwidth}{!}{\includegraphics*{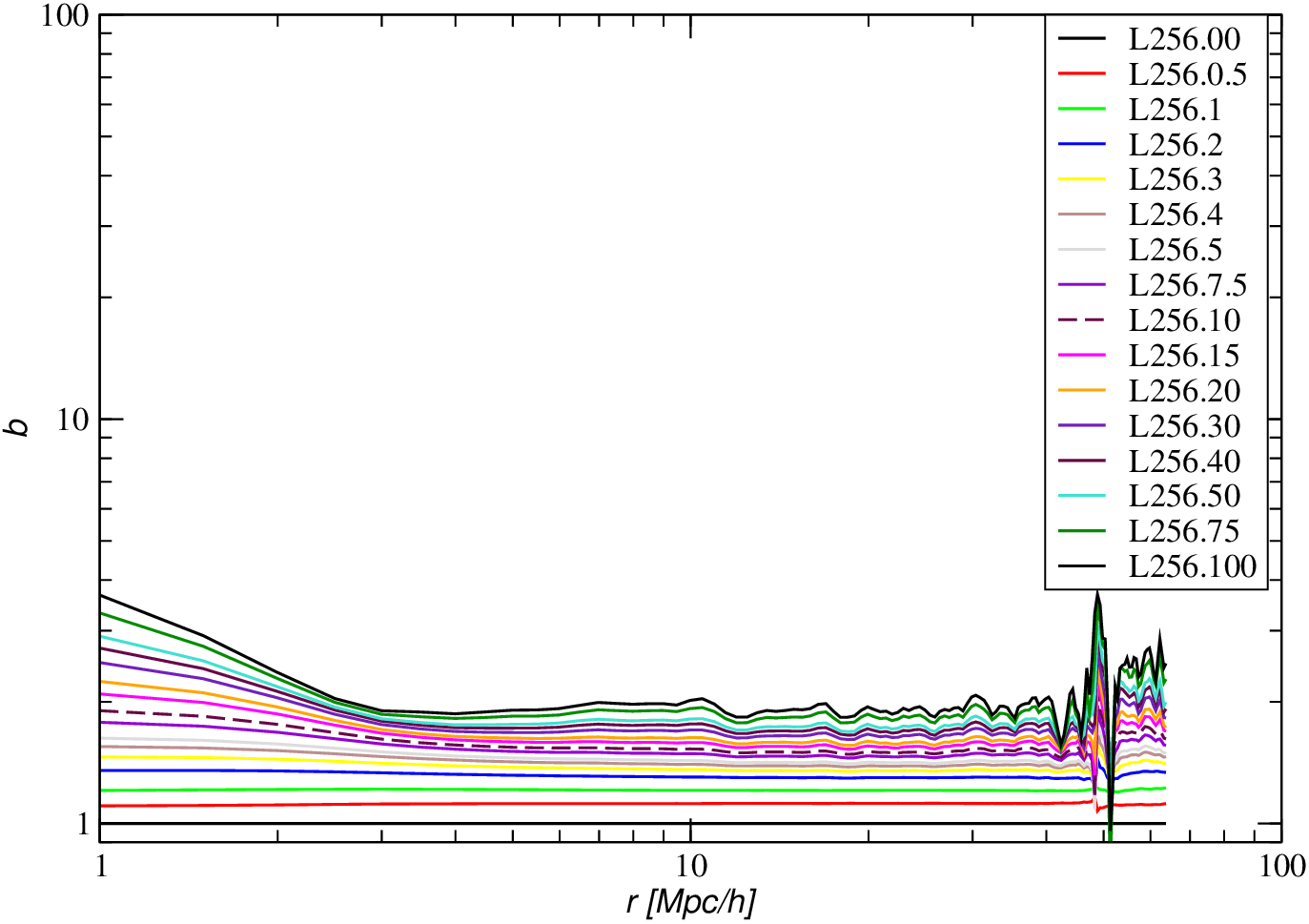}}
\hspace{1mm}  
\resizebox{0.32\textwidth}{!}{\includegraphics*{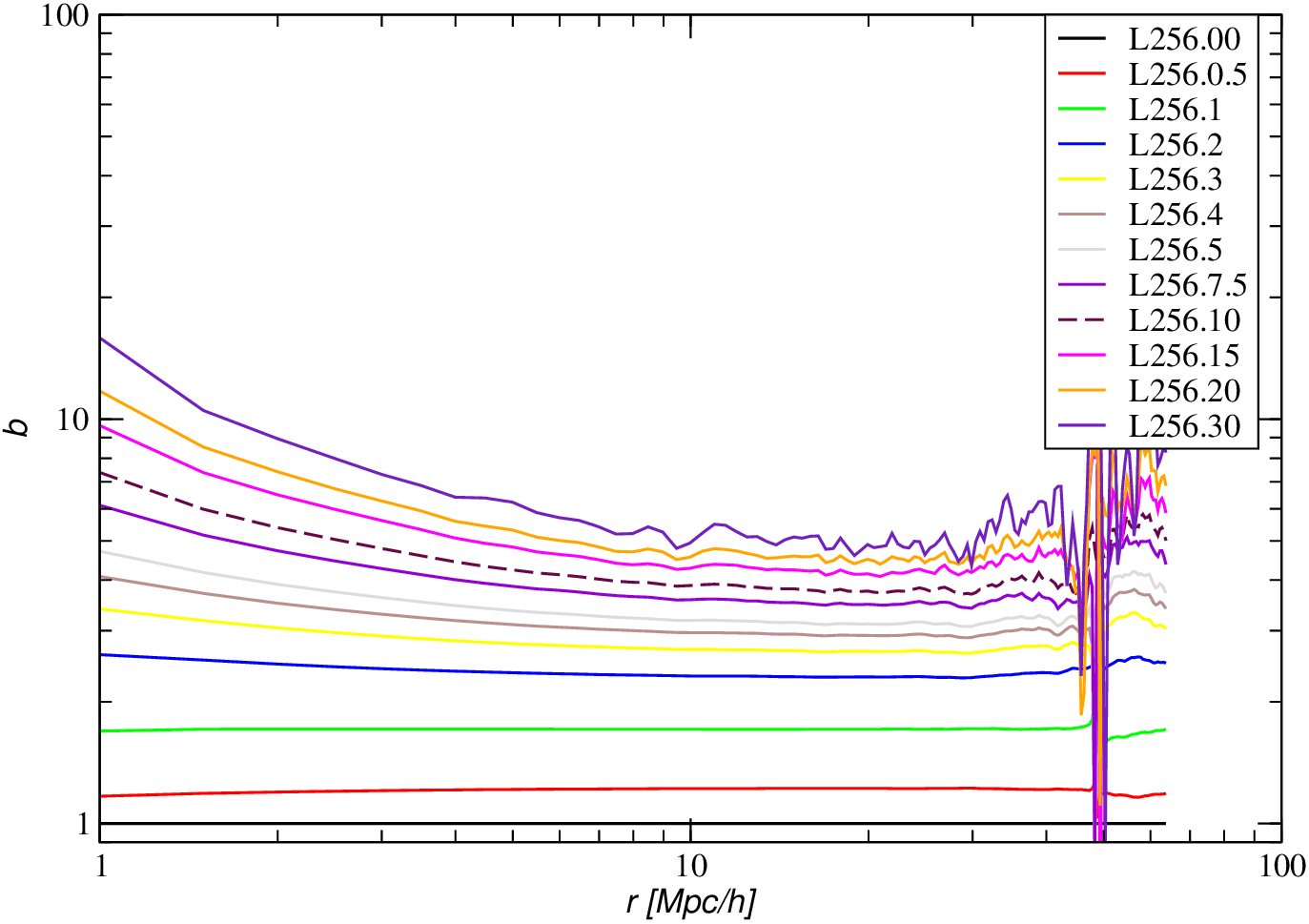}}
\hspace{1mm}  
\resizebox{0.32\textwidth}{!}{\includegraphics*{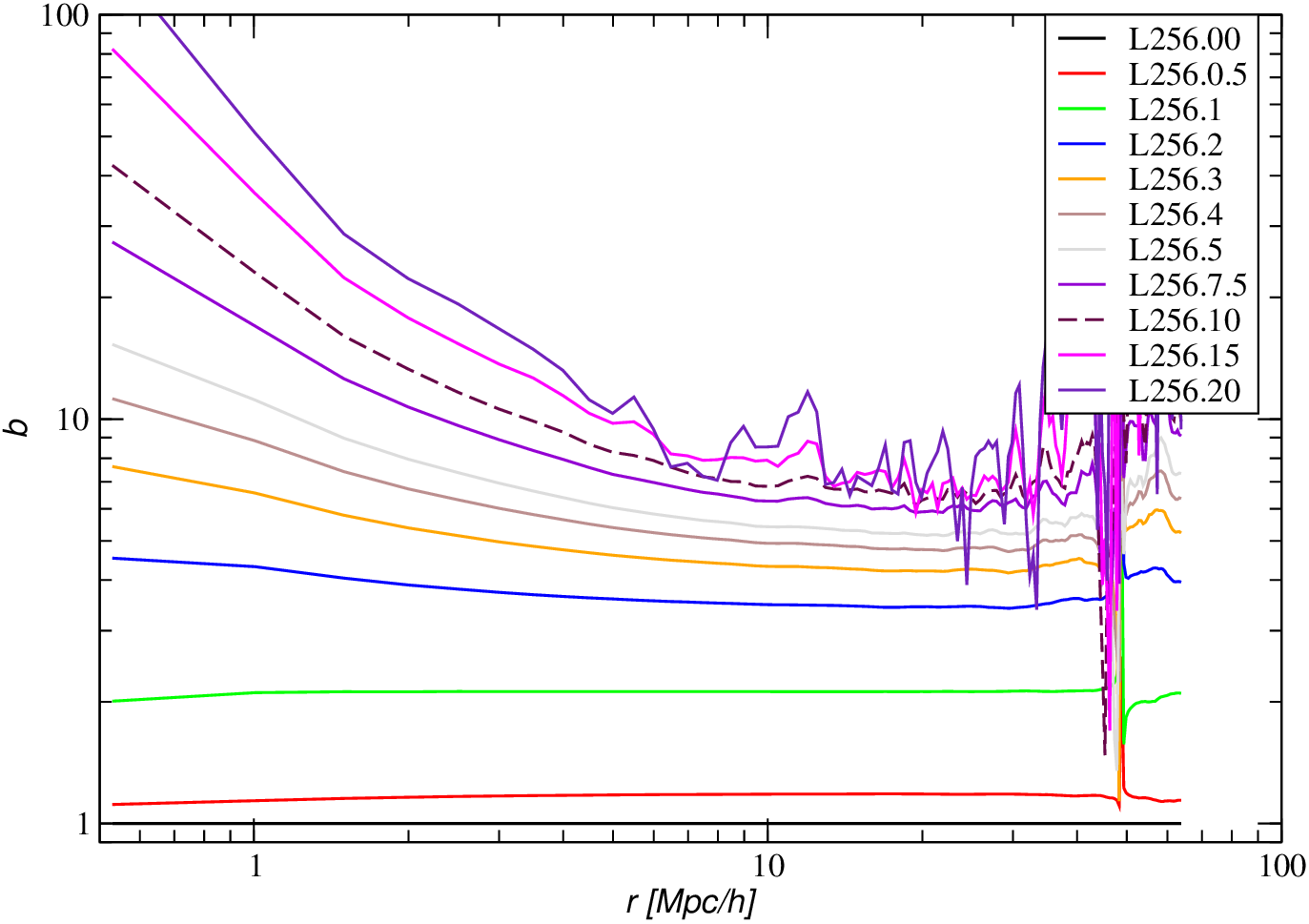}}\\
\hspace{1mm}  
\resizebox{0.32\textwidth}{!}{\includegraphics*{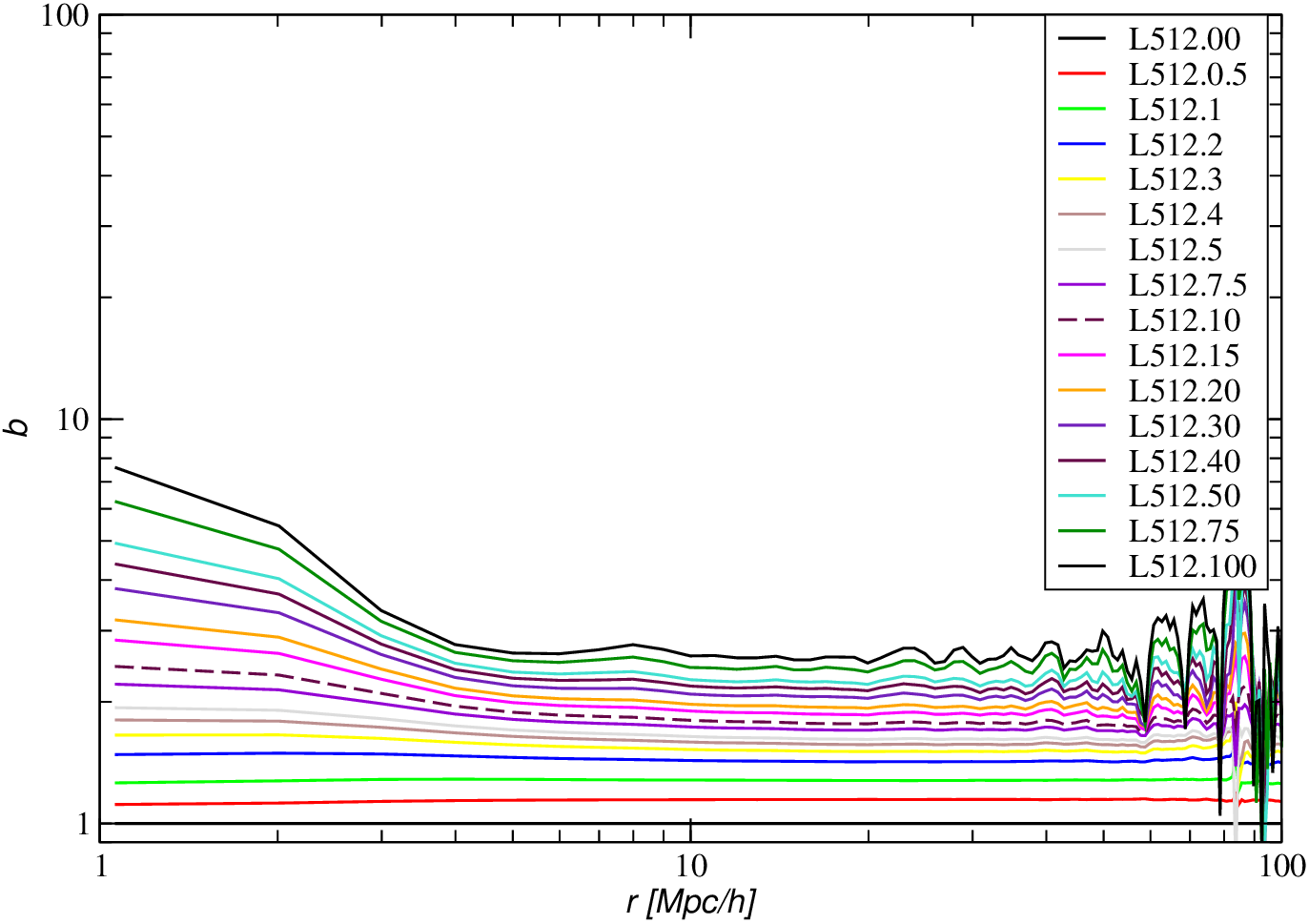}}
\hspace{1mm}  
\resizebox{0.32\textwidth}{!}{\includegraphics*{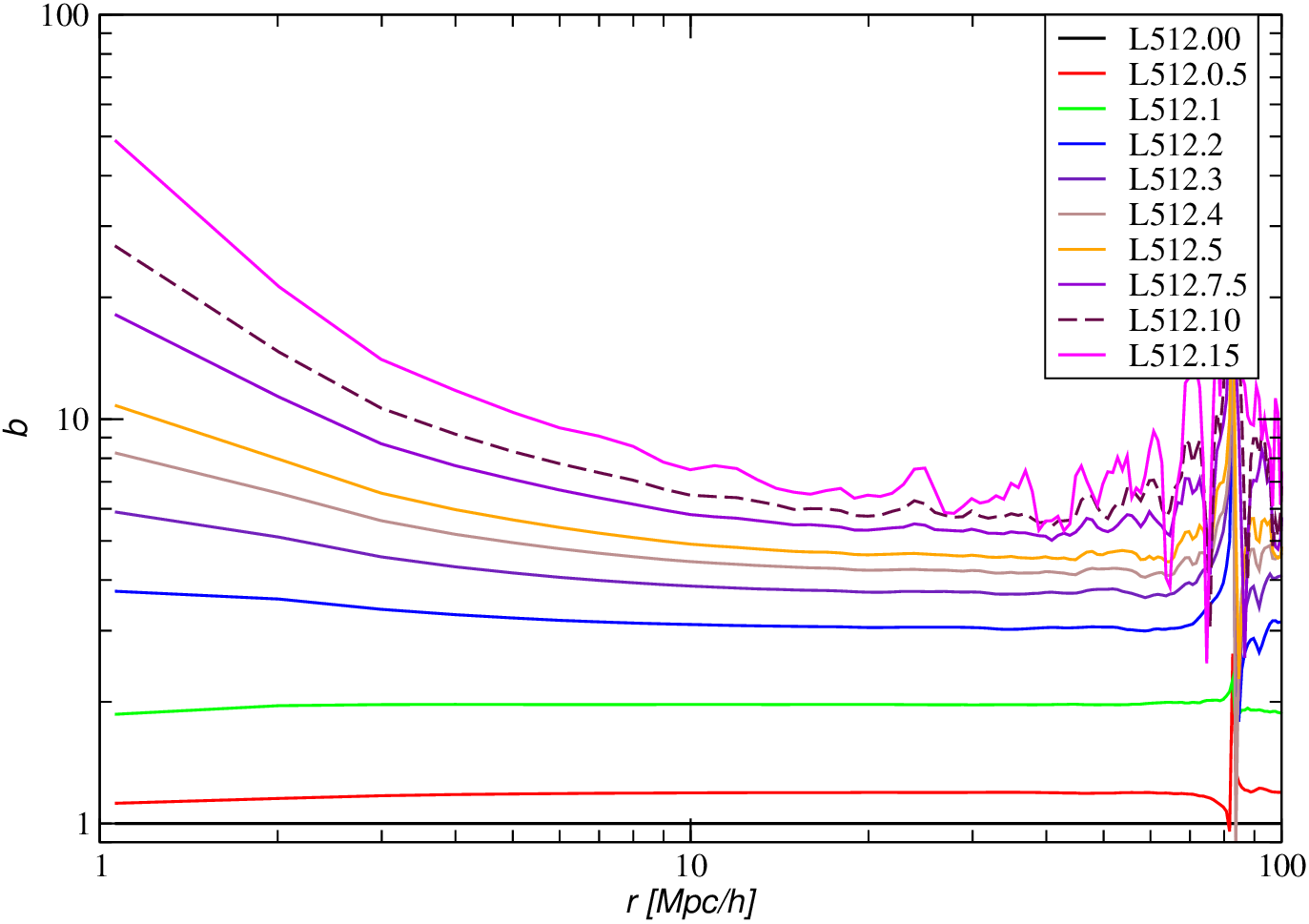}}
\hspace{1mm}  
\resizebox{0.32\textwidth}{!}{\includegraphics*{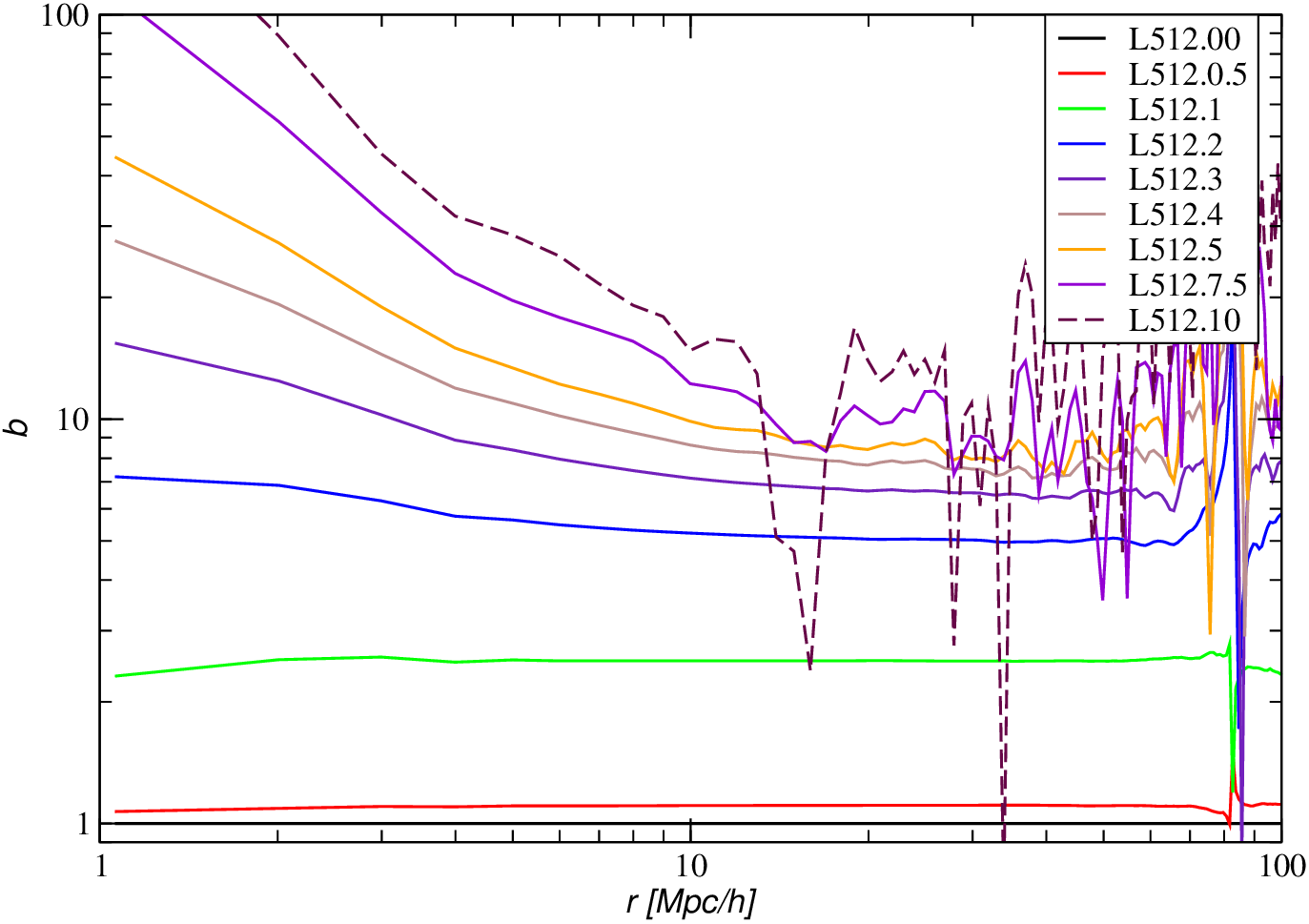}}\\
\hspace{1mm}  
\resizebox{0.32\textwidth}{!}{\includegraphics*{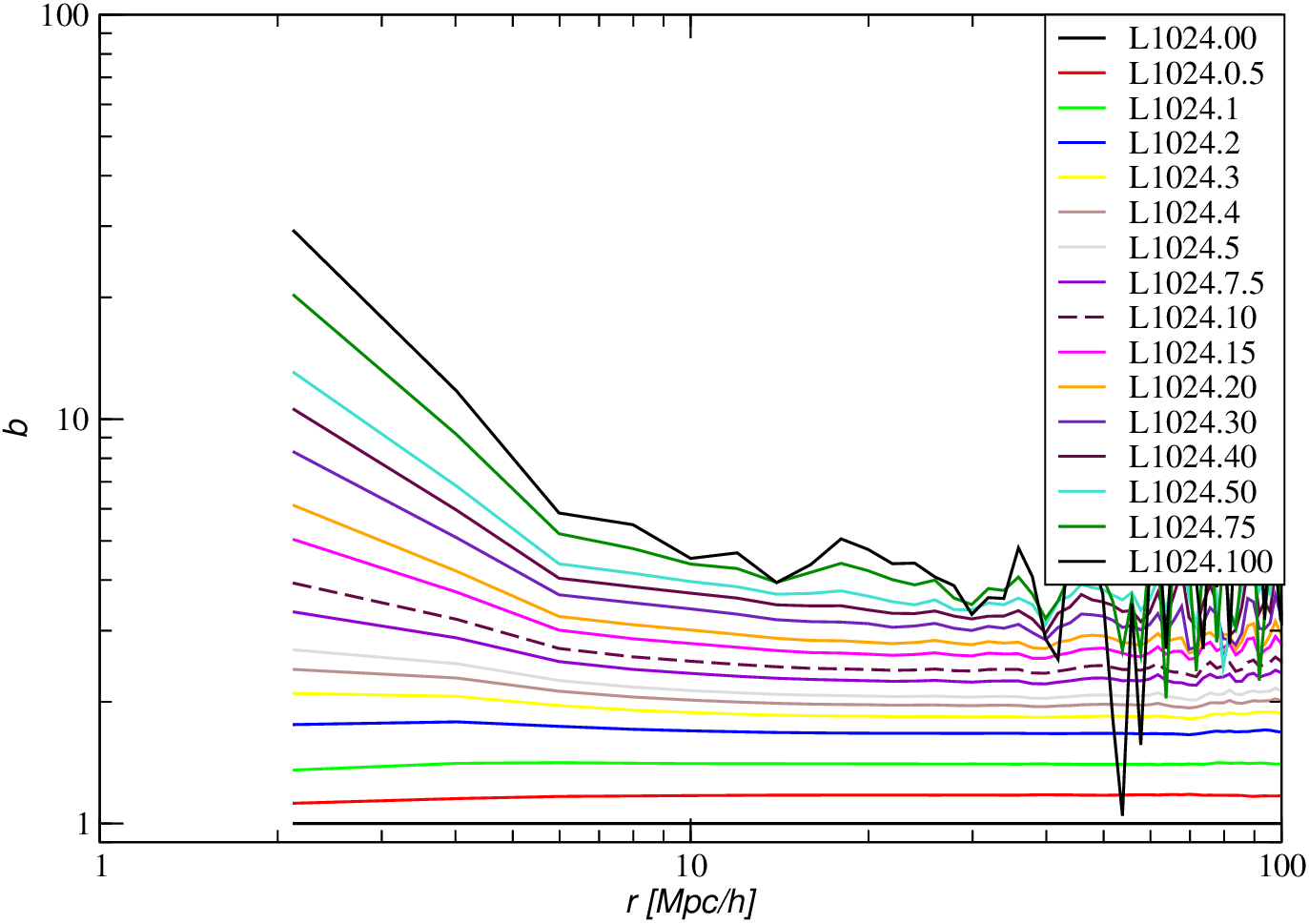}}
\hspace{1mm}  
\resizebox{0.32\textwidth}{!}{\includegraphics*{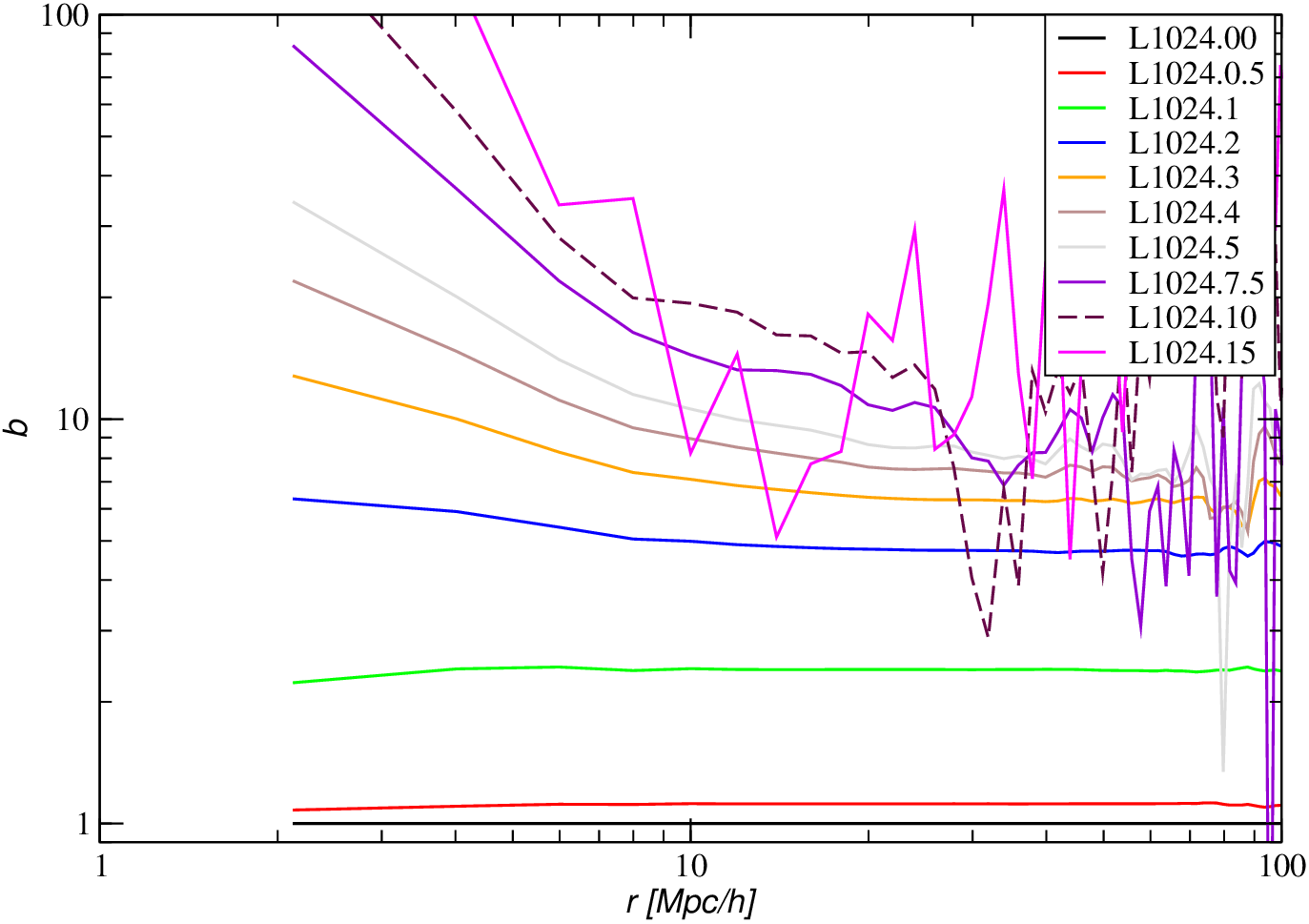}}
\hspace{1mm}  
\resizebox{0.32\textwidth}{!}{\includegraphics*{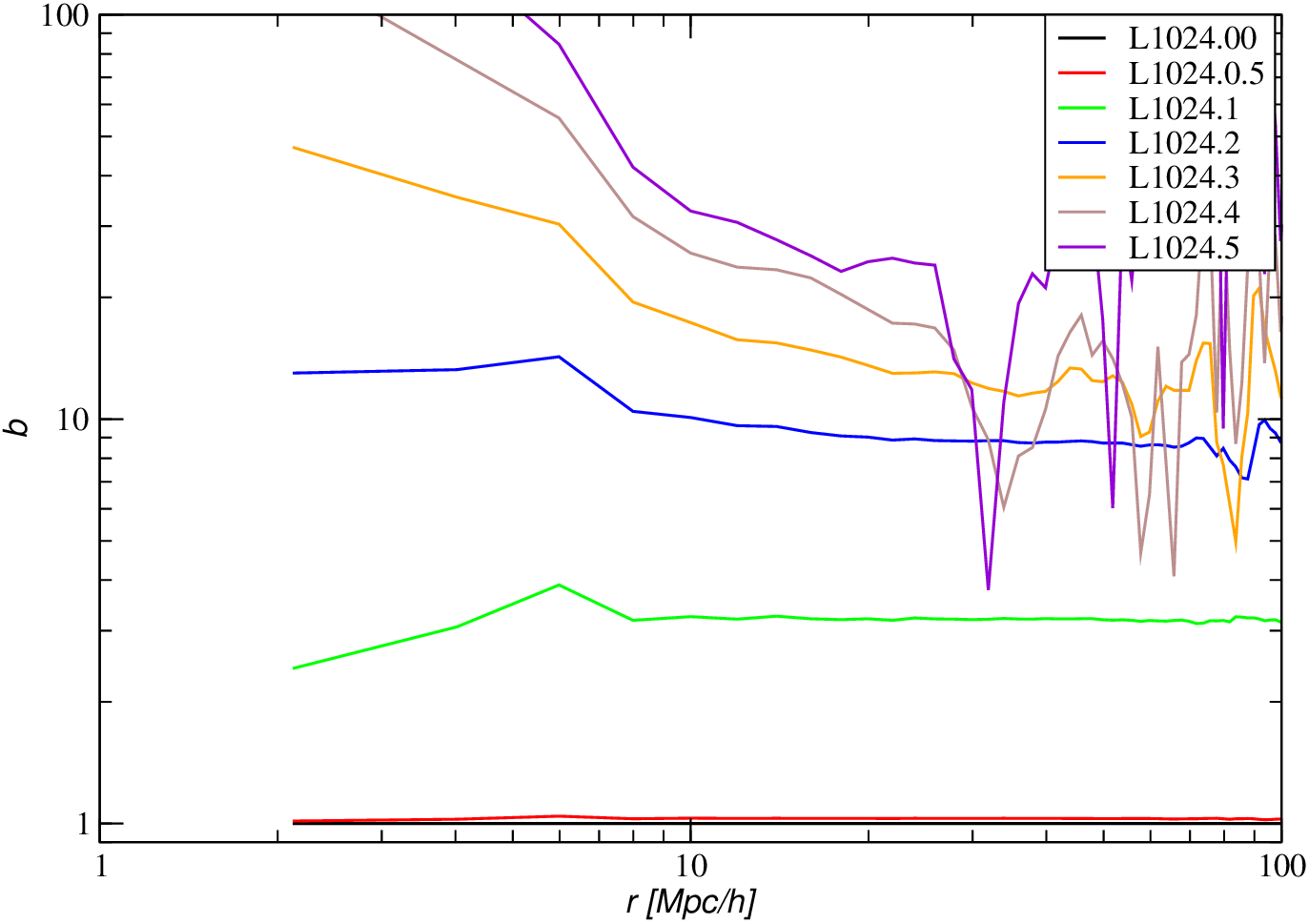}}\\
\caption{Bias functions, $b(r)$, as functions of the galaxy pair
  separation $r$ for epochs $z=0,~3,~5$, shown respectively in the
  left, central and right panels.  Upper panels show bias of  the model L256,
  middle panels  of  the model L512, and lower panels  of  the L1024 model.
  Density limits $\rho_0$ are indicated as symbol labels. }
\label{fig:Fig5} 
\end{figure*}

We show in Fig.~\ref{fig:Fig4} a series of correlation functions for
various particle density limits $\rho_0$.  Left, central and right
panels are for epochs $z=0,~3,~5$, respectively,.  Results for L256,
L512 and L1024 models are in upper, middle and lower panels, all found
with grid resolution $N_{\mathrm{grid}}=1024^3$. Here samples with
density limit $\rho_0=0$ are the samples with all particles and
represents the whole DM simulation. On smaller separations
$r \le 5~\Mpc$ correlation functions describe the
distribution of particles in DM halos, for a discussion of this
phenomenon see \citet{Einasto:2020aa}. For larger separations
correlation functions describe fractal properties of the cosmic web.
The fractal dimension function, $D(x)= 3+\gamma(r)$, is defined
through the logarithmic gradient of the correlation function,
$\gamma(r)= d \log g(r)/ d \log r$, where $g(r)=1 +\xi(r)$, see
\citet{Einasto:2020aa}.

Central and right panels of Fig.~\ref{fig:Fig4} present correlation
functions for epochs $z=3,~5$ of simulations.  The Figure shows that
reliable correlation functions can be found for these epochs only in a
limited range of particle density limits, $\rho_0$ up to
$\rho_0\approx 30$. This limit is due to the sparsity of particles
with density labels above these limits at respective redshifts, 
see Fig.~\ref{fig:Fig2} for the cumulative distribution of
local densities of particles.

\subsection{Evolution of bias functions of $\Lambda$CDM models}

In Fig.~\ref{fig:Fig5} we present bias functions, Eq.~(\ref{bias}),
for epochs $z=0,~3,~5$.  Upper, middle and lower panels present bias
functions for simulations L256, L512 and L1024, respectively, found
with grid resolution $N_{\mathrm{grid}}=1024^3$.  As noted above, bias
functions have a plateau at $r \geq 6~\Mpc$ for the present epoch
$z=0$, see Fig.~\ref{fig:Fig5}.  We used this plateau to measure
relative amplitudes of the correlation functions, which define bias
parameters.  In this separation range, bias functions change,
therefore the location of the reference point influences our
results. Following \citet{Einasto:2021ti} we used initially the separation
$r_0=r_{6}=6~\Mpc$.

Higher amplitudes of bias functions at small separations are due to
the influence of halos.  The comparison of panels for different
simulations shows that the influence of halos is limited to separation
$r \approx 5~\Mpc$ for the present epoch $z=0$, and $r\approx 10~\Mpc$
for earlier epochs of the L256 simulation.  For L512 and L1024
simulations the influence of halos is seen up to higher separations.
To avoid the influence of halos we accepted  in the final analysis 
a higher $r_0$ value to find  bias parameters, $r_0=r_{10}=10~\Mpc$.

\subsection{Evolution of bias parameters with cosmic epoch}

\begin{figure*}
\centering 
\hspace{1mm}  
\resizebox{0.32\textwidth}{!}{\includegraphics*{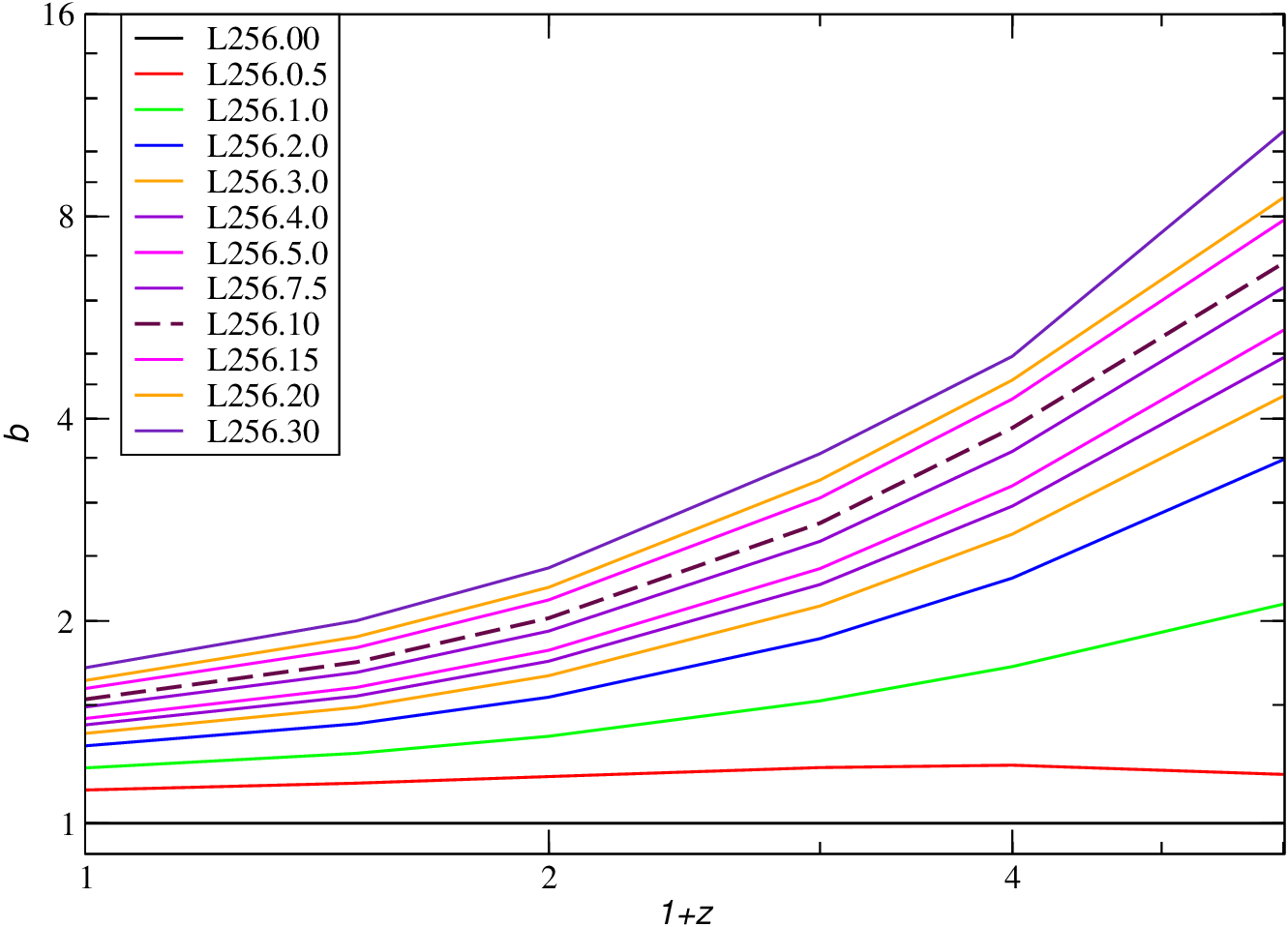}}
\hspace{1mm}  
\resizebox{0.32\textwidth}{!}{\includegraphics*{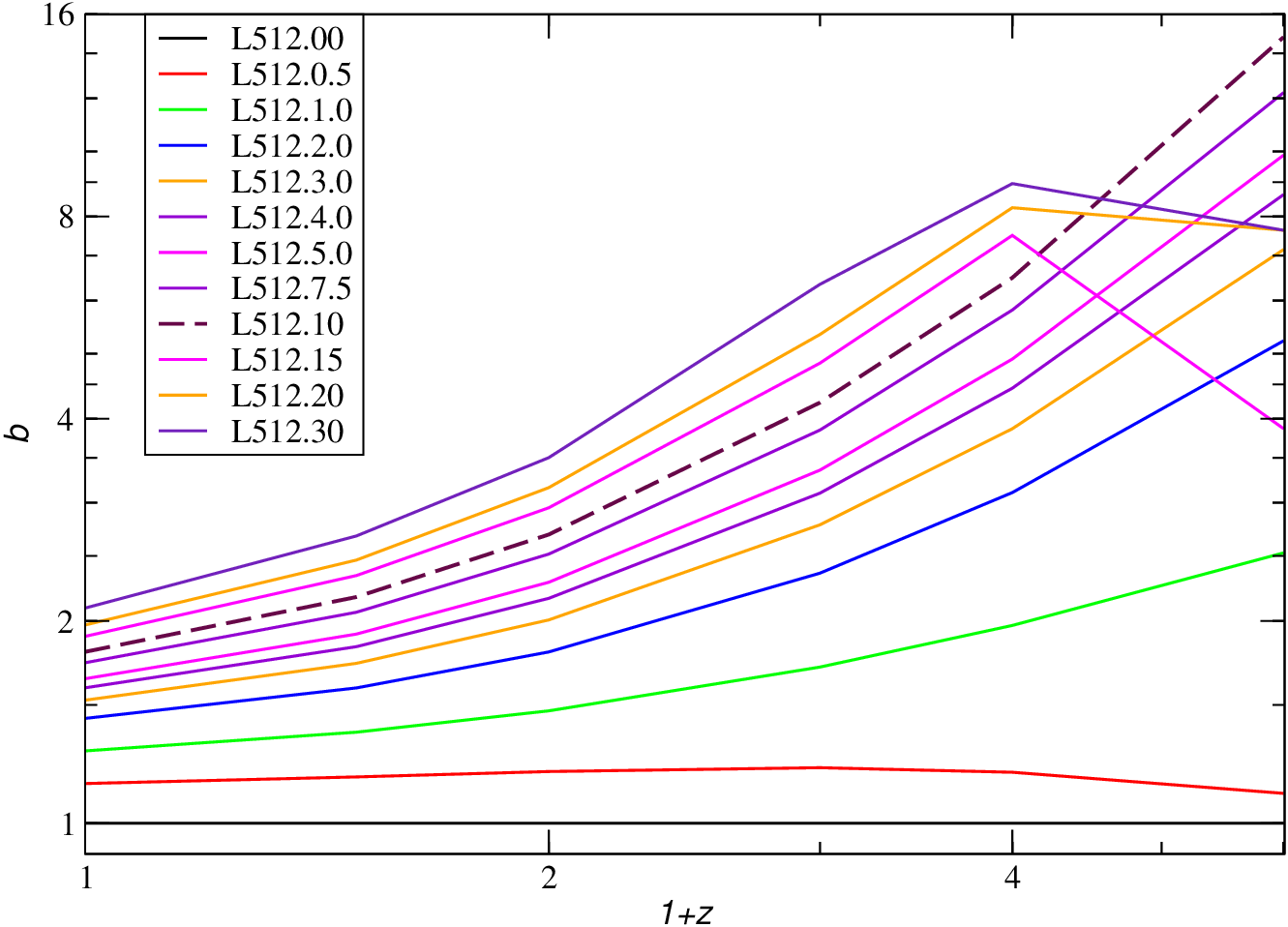}}
\hspace{1mm}  
\resizebox{0.32\textwidth}{!}{\includegraphics*{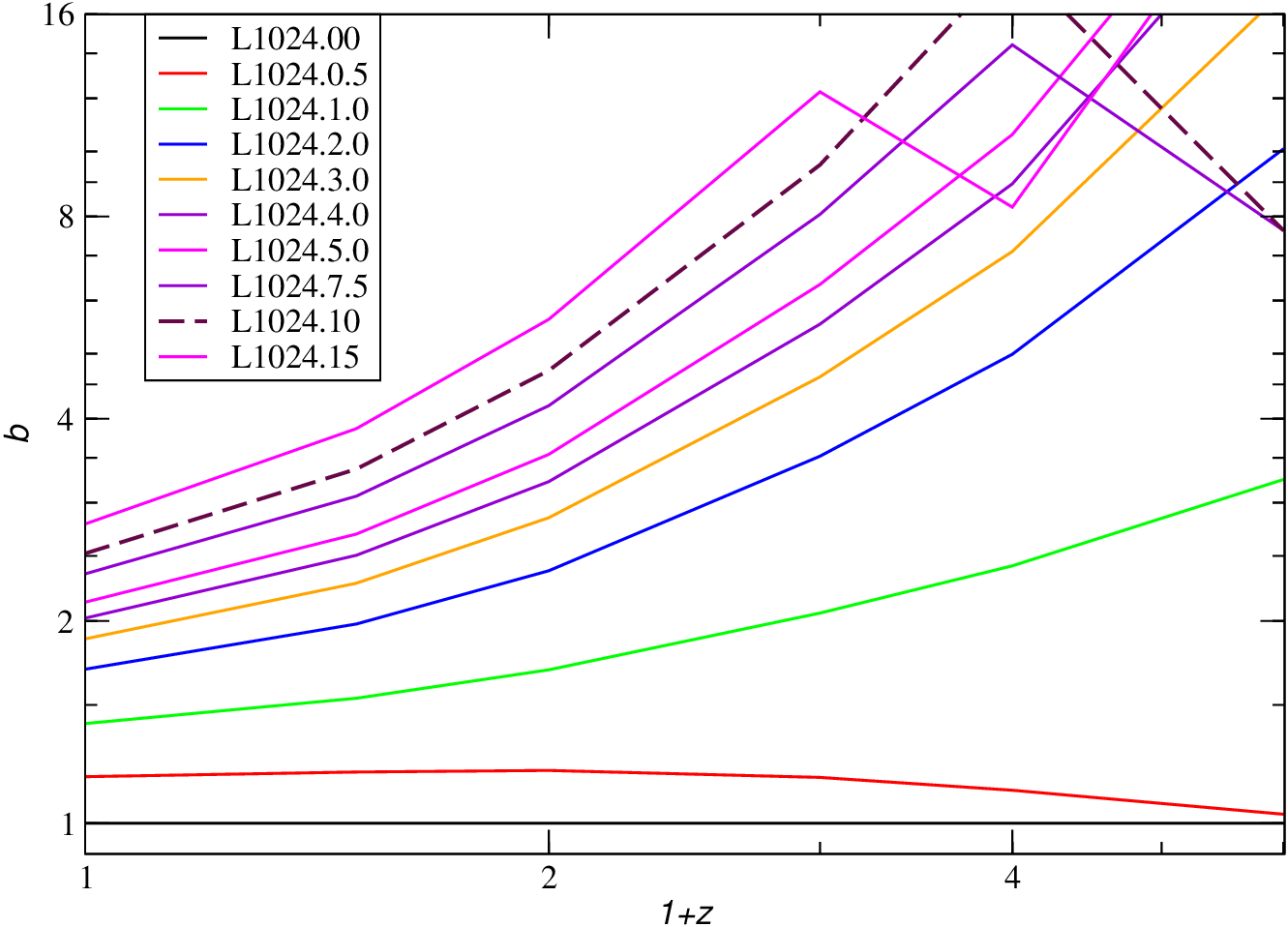}}\\
\hspace{1mm}  
\resizebox{0.32\textwidth}{!}{\includegraphics*{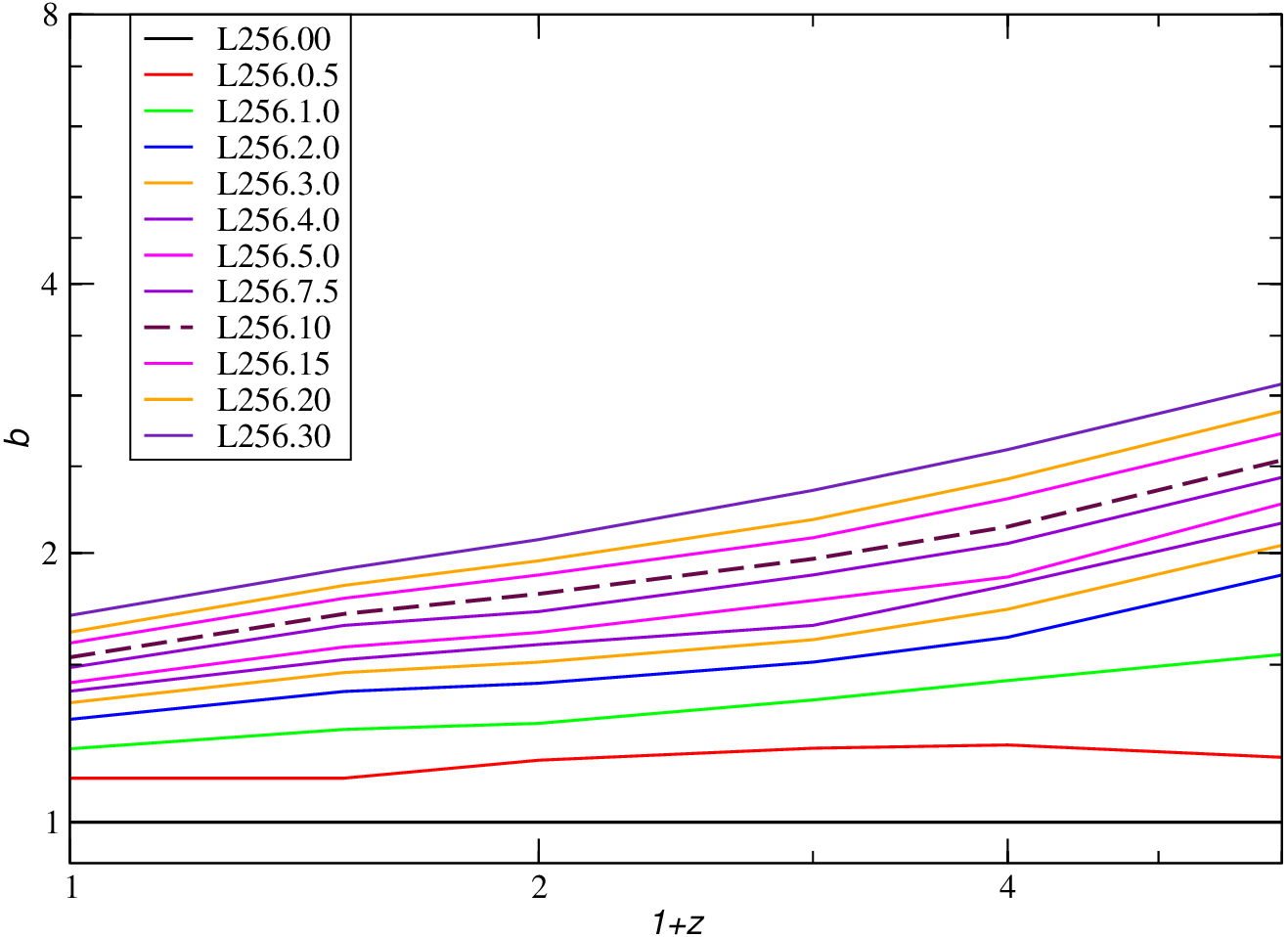}}
\hspace{1mm}  
\resizebox{0.32\textwidth}{!}{\includegraphics*{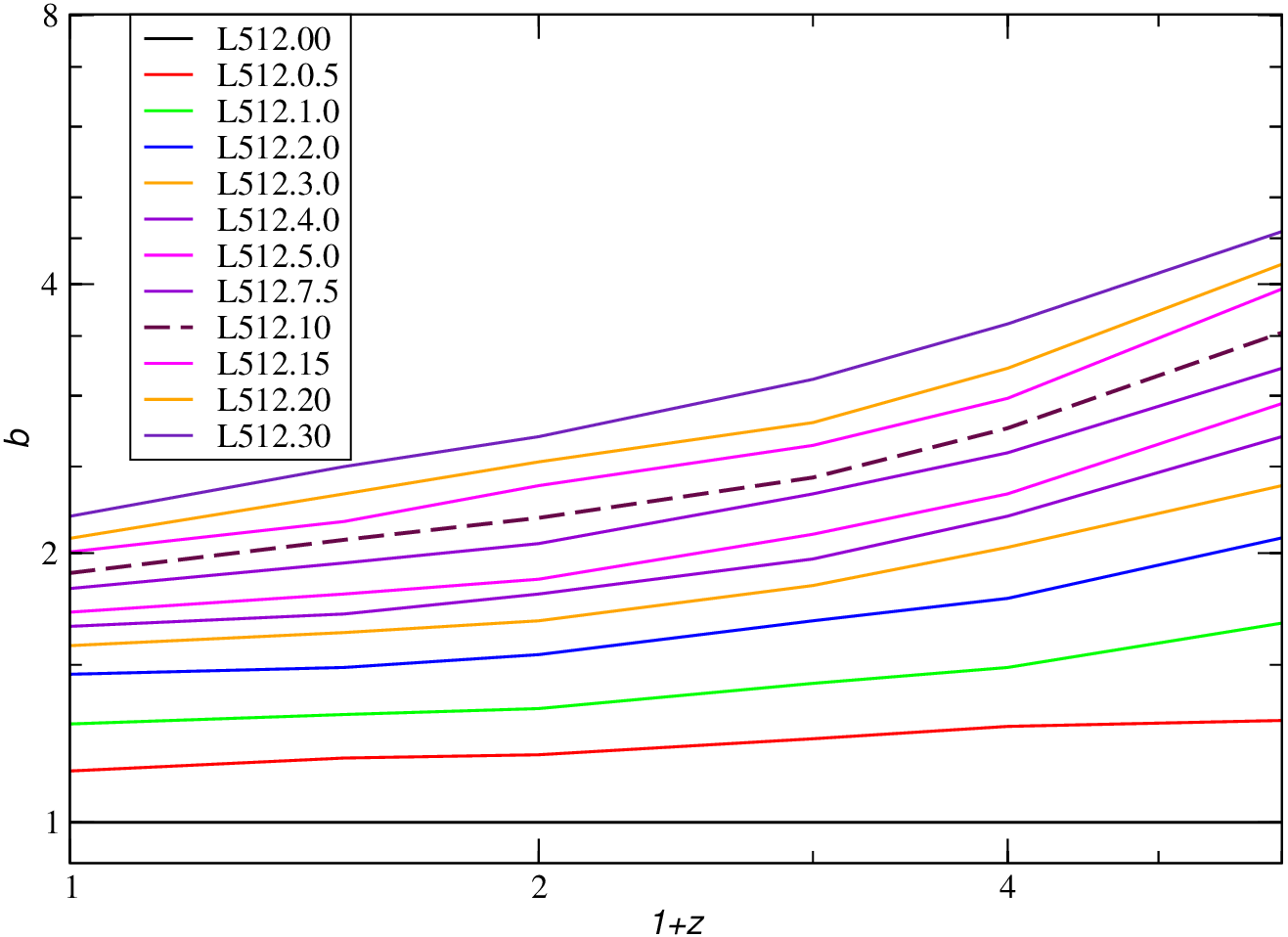}}
\hspace{1mm}  
\resizebox{0.32\textwidth}{!}{\includegraphics*{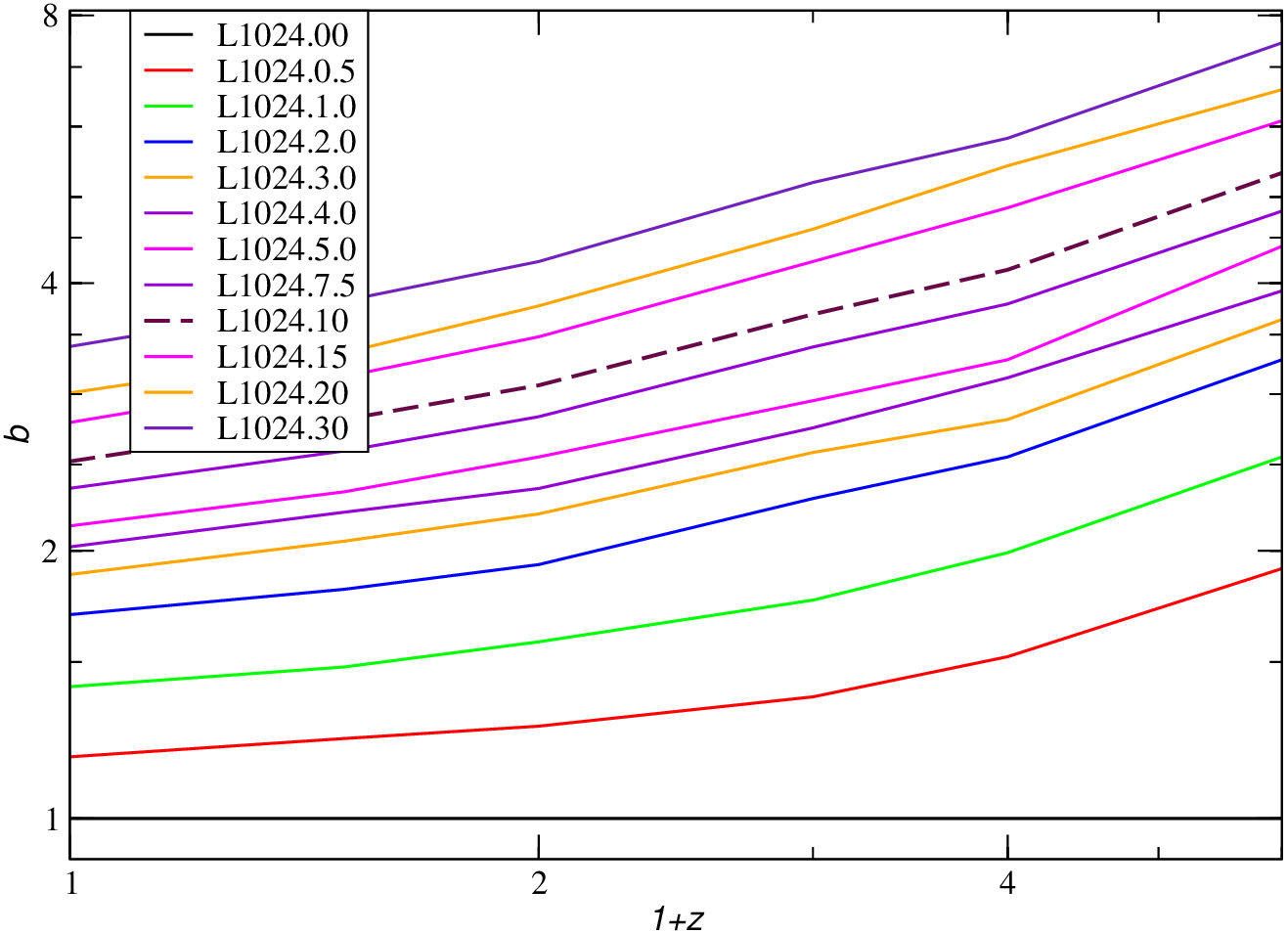}}
\caption{Top: Evolution of bias parameter values with epoch of
  simulations $z$, using fixed density limits $\rho_0$, indicated as
  symbol labels.  Correlation and bias functions were found with
  resolution $N_{\mathrm{grid}}=1024^3$.  Bottom: Evolution of bias
  parameters with epoch of simulations $z$.  Bias parameters are
  calculated for identical fraction of particles, separately for
  different simulations and epochs.  Density limits $\rho_0$ are
  indicated as symbol labels.  }
\label{fig:Fig6} 
\end{figure*}

Upper panels of Fig.~\ref{fig:Fig6} present the evolution of bias
parameters with redshift $z$ for models L256, L512 and L1024, found
for a series of particle density limits, $\rho_0$, shown as model name
second index.  Here we used grid resolution $N_{\mathrm{grid}}=1024^3$
and separation $r_{10}=10~\Mpc$.  Fig.~\ref{fig:Fig5} shows that bias
functions for particle density limits $\rho_0 >5$ have wiggles already
at the separation $r \approx 10~\Mpc$, used in the determination of
bias parameters. These wiggles are due to decreasing number of
particles at respective separations, and decrease bias parameter
values for early epochs and high particle density limit $\rho_0$ of
models L512 and L1024. However, in the further analysis with reduced
bias parameters we do not use these regions of bias functions.

Lower panels of Fig. ~\ref{fig:Fig6} and Tables \ref{Tab2} to
\ref{Tab4} present bias parameters as functions of the cosmic epoch
$z$, calculated for identical fraction of particles at all simulations
and epochs, as described in the next subsection, and using separation
$r_0=R_{10}=10~\Mpc$ to find bias parameters.  As expected, the
reduced bias parameter values are lower than values, found for fixed
particle density levels.  The comparison of bias parameters of models
L256, L512 and L1024 shows, that bias parameters for models with
larger simulation box sizes are higher than in the model of smaller
size.  We discuss this effect in more detail in the next Section.

{\scriptsize 
 \begin{table*}
\caption{Bias parameters of L256 particle-density-limited models} 
\label{Tab2}                         
\centering
\begin{tabular}{lrrrrrrrr}
\hline  \hline
Sample   & $F(\rho_0)$& $1/F$&$z=0$& $z=0.5$& $z=1$ &$z= 2$& $z=3$& $z=5$\\  
\hline  
(1)&(2)&(3)&(4)&(5)&(6)&(7)&(8)&(9)\\ 
\hline  
L256.00  &  1.00000 & 1.0000&1.000  &   1.000  &   1.000  &  1.000  &   1.000  &  1.000\\
L256.05 &  0.89665 & 1.1153 &   1.120   &  1.147  &   1.173  &   1.21&    1.220  &   1.182\\
L256.1   & 0.81224 & 1.2312  &     1.208 &   1.27&   1.29&   1.37&   1.44&   1.54\\
L256.2   & 0.71285 &  1.4028 &     1.303 &   1.40&   1.43&   1.51&   1.61&   1.89\\
L256.3   & 0.65022 &  1.5379  &     1.360&   1.47&   1.51&   1.60&   1.73&   2.04\\
L256.4   &  0.60490 & 1.6532   &     1.401 &   1.52&   1.58&   1.66&   1.84&   2.16\\
L256.5   & 0.56995 & 1.7545   &    1.432 &   1.57&   1.63&   1.77&   1.88&   2.27\\
L256.7.5& 0.50768 & 1.9697   &   1.489 &   1.66&   1.72&   1.89&   2.05&   2.43\\
L256.10 &0.46503 &  2.1504  &   1.529 &   1.71&   1.80&   1.97&   2.14&   2.54\\
L256.15 & 0.40695 & 2.4573  &   1.586 &   1.78&   1.89&   2.08&   2.30&   2.72\\
L256.20 &0.36648 &  2.7287  &  1.631 &   1.84&   1.96&   2.18&   2.42&   2.88\\
L256.30 &0.30907 &  3.2355  &   1.703 &   1.92&   2.07&   2.35&   2.61&   3.09\\
\hline 
\end{tabular} \\
{
The columns show the
(1) sample name; 
(2)  the fraction of particles $F(\rho_0)$ at particle-density limit
$\rho_0$;
(3) $1/F(\rho_0)$:
(4) - (9) the bias parameters, calculated for epochs $z=0,~0.5,~1,~2,~3,~5$.
}
\end{table*} 
}

{\scriptsize 
\begin{table*}
\caption{Bias parameters of L512 particle-density-limited models} 
\label{Tab3}                     
\centering
\begin{tabular}{lcrrrrrrr}
\hline  \hline
Sample   & $F(\rho_0)$&$1/F(\rho_0)$&$z=0$& $z=0.5$& $z=1$ & $z=2$& $z=3$& $z=5$ \\  
\hline  
(1)&(2)&(3)&(4)&(5)&(6)&(7)&(8)&(9)\\ 
\hline  
L512.00   & 1.00000& 1.000&1.000  &   1.00  &   1.00   &  1.00  &   1.00   &  1.00\\
L512.0.5 &  0.89972 & 1.112&1.141   &    1.18  & 1.19 & 1.24  & 1.28  &  1.30\\
L512.1   &  0.79682 & 1.255&1.288   &     1.32  & 1.34  & 1.43 &  1.49  &  1.67\\
L512.2   &  0.67842 & 1.474&1.464   &      1.49 &  1.54  & 1.68  & 1.78  &  2.08\\
L512.3   &  0.60629 & 1.649&1.576   &     1.63  &  1.68 & 1.84 &  2.03 &  2.38\\
L512.4   &  0.55536  & 1.801& 1.656  &    1.71   & 1.80 & 1.97 & 2.20 &  2.70\\
L512.5   &  0.51659 & 1.936& 1.718   &     1.80  &  1.87 & 2.10 &  2.33 &   2.94\\
L512.7.5 &  0.44900 & 2.227& 1.826  &      1.95 &  2.05 & 2.33 &  2.59  &  3.22\\
L512.10 &  0.40359 & 2.478&1.900  &     2.07 &   2.19 & 2.43 &  2.76  &  3.53\\
L512.15 &  0.34314 & 2.914&2.006   &    2.17 &   2.38 & 2.64 &  2.98 &   3.95\\
L512.20 &  0.30235 & 3.307&2.078   &   2.33  &  2.53 & 2.80 &  3.22  &  4.21\\
L512.30 &  0.24726 & 4.044&2.199   &   2.50  &  2.70 & 3.13  & 3.61  &  4.58\\
\end{tabular} \\
{
The columns show the
(1) sample name; 
(2)  the fraction of particles $F_0$ at particle-density limit
$\rho_0$;
(3) parameter $b_c=1/F_0$;
(4) - (9) the bias parameters, calculated for epochs $z=0,~0.5,~1,~2,~3,~5$.
}
\end{table*} 
}

{\scriptsize 
\begin{table*}
\caption{Bias parameters of L1024 particle-density-limited models} 
\label{Tab4}                     
\centering
\begin{tabular}{lcrrrrrrr}
\hline  \hline
Sample   & $F(\rho_0)$&$1/F(\rho_0)$&$z=0$& $z=0.5$& $z=1$ & $z=2$& $z=3$& $z=5$ \\  
\hline  
(1)&(2)&(3)&(4)&(5)&(6)&(7)&(8)&(9)\\ 
\hline  
L1024.00   & 1.00000& 1.000&1.000  &   1.00  &   1.00   &  1.00  &   1.00   &  1.00\\
L1024.0.5 &  0.90150& 1.109&1.173 &  1.23 &   1.27 &   1.37 &   1.52 &   1.91\\
L1024.1   &  0.75359& 1.327&1.407 &  1.48 &   1.58 &   1.76 &   1.99 &   2.55\\
L1024.2   &  0.58092& 1.721&1.695 &  1.81 &   1.93 &   2.29 &   2.55 &   3.28\\
L1024.3   & 0.48154 & 2.077&1.881 &  2.05 &   2.20 &   2.58 &   2.81 &   3.64\\                            
L1024.4   &  0.41490  & 2.410&2.020 &  2.21 &   2.35 &   2.75 &   3.13 &   3.92\\
L1024.5   & 0.36639 & 2.729& 2.133 &  2.33 &   2.55 &   2.95 &   3.28 &   4.40\\
L1024.7.5 & 0.28614 & 3.495&2.351 &  2.59 &   2.83 &   3.39 &   3.79 &   4.82\\
L1024.10 & 0.23579 & 4.241&2.521 &  2.81 &   3.07 &   3.69 &   4.14 &   5.32\\
L1024.15 &  0.16869& 5.928&2.788 &  3.14 &   3.48 &   4.23 &   4.86 &   6.09\\
L1024.20 & 0.13631 & 7.336&3.010 &  3.34 &   3.77 &   4.60 &   5.42 &   6.60\\
L1024.30 &  0.09132&10.951 &3.394 &  3.82 &   4.23 &   5.19 &   5.82 &   7.45\\
\end{tabular} \\
{
The columns show the
(1) sample name; 
(2)  the fraction of particles $F_0$ at particle-density limit
$\rho_0$;
(3) parameter $b_c=1/F_0$;
(4) - (9) the bias parameters, calculated for epochs $z=0,~0.5,~1,~2,~3,~5$.
}
\end{table*} 
}

\begin{figure*}
\centering 
\hspace{1mm}  
\resizebox{0.35\textwidth}{!}{\includegraphics*{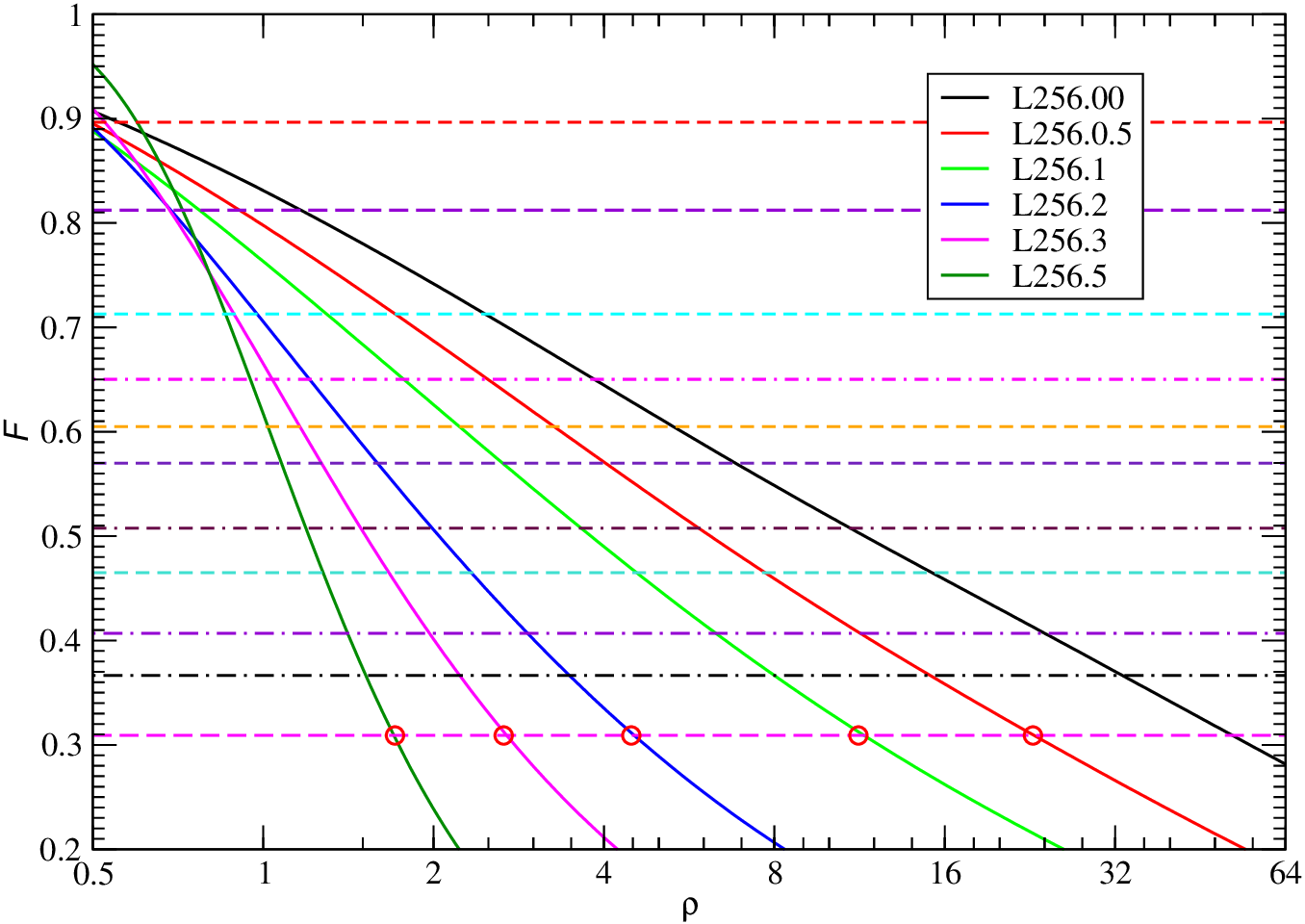}}
\hspace{1mm}  
\resizebox{0.35\textwidth}{!}{\includegraphics*{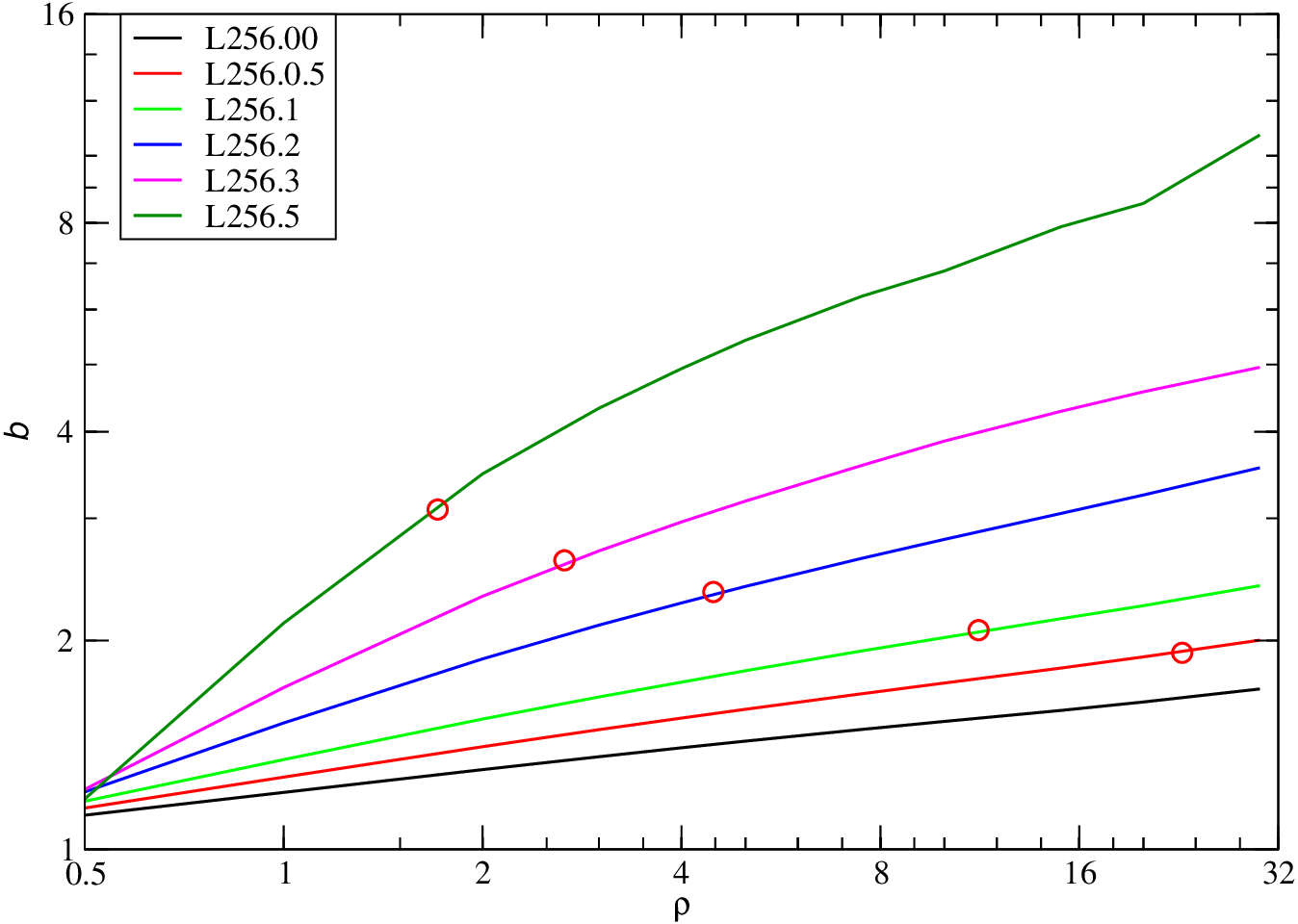}}
\caption{Left: Particle density distribution of model L256 for epochs
  $z=0,~0.5,~1,~2,~3,~5,~10$.  Horisontal lines show cumulative
  density levels used in calculation of bias values.  Right: Bias
  values for various particle density levels $\rho_0$ of the model
  L256 for the same epochs. }
\label{fig:Fig7} 
\end{figure*} 

\subsection{Reduction of bias parameters to identical fraction  of
  galaxies}

In the first stage of our study we used identical particle density
limits $\rho_0$ to select biased samples of particles for all
simulation epochs.  However, particle densities evolve with time, and
identical particle density limits correspond at different epochs to
different objects.  To get bias parameters for objects of comparable
properties it is needed to bring bias values to a comparable system.
To do this we used identical fraction of particles, separately for
each model and simulation epoch.

This reduction was made in a two step procedure. First we found the
cumulative distribution of the fraction of particles
$F(\rho_0)= N(\rho_0)/N(0)$ as function of $\rho_0$, as done in
Fig.~\ref{fig:Fig3}. Left panel of Fig.~\ref{fig:Fig7} shows these
distributions for the simulation L256 and epochs
$z=0,~0.5,~1,~2,~3,~5$.  Horisontal dashed lines show fraction levels,
used in the model L256 at present epoch $z=0$ to select particles for
biased sample populations for particle density limits $\rho_0=0.5$ to
$\rho_0=30$.  These fractions are given in Table \ref{Tab2} for the
L256 simulation, and in Tables \ref{Tab3} and \ref{Tab4} for L512 and
L1024 simulations.  Using these $F(\rho_0)$ values we found by
interpolation for each evolution epoch $z$ and the fraction
$F(\rho_0)$ the particle density $\rho_0$, which at given epoch
corresponds to the same population, defined by the respective
fraction. For the L256.30 sample and simulation epochs
$z=0.5,~1.0,~2.0,~3.0,~5.0$ particle density $\rho_0$ values,
corresponding to $F(\rho_0)= 0.30907$, are shown as red circles.

In the second step we used $\rho_0$ values, obtained in the first step
at the same fraction of particles $F(\rho_0)$, to find bias $b$ values
for corresponding epochs, using the $b(\rho_0)$ diagram, shown in
the right panel of Fig.~\ref{fig:Fig7}.  Again points in the $b(\rho_0)$
diagram, corresponding to the L256.30 sample,  are shown as  red
circles. In both panels points corresponding to identical epochs are
plotted with   curves of identical colour. 
This interpolation procedure was applied to found reduced bias values
for simulations L256, L512 and L1024, given  in Tables \ref{Tab2},
\ref{Tab3} and \ref{Tab4} by two decimal digits.

This procedure does not take into account the increase of masses of
galaxies with time. Actually masses of halos and galaxies 
increase with time \citep{Chiang:2013tq}.  However, this increase is
rather modest \citep{Park:2022va}, and will
not change our results considerably, see the Discussion.

\subsection{Error analysis}

The number of particles in simulations is very high ($512^3$), thus
random errors of correlation and bias functions are very small, as
found by \citet{Einasto:2021ti}. Only at early epochs the number of
particles for high $\rho_0$ values is small and errors are large, as
seen from Figs.~\ref{fig:Fig4} and \ref{fig:Fig5}.  However, these
regions of bias function are not used for further analysis with
reduced $\rho_0$ values.  In the $b(\rho_0)$ diagram in the right
panel of Fig.~\ref{fig:Fig7} red coloured points are for samples with
particle density limit $\rho_0=30$. All points for samples with lower
particle density limit are situated at lower reduced $\rho_0$ levels,
where random errors are small.
Thus essential  errors of our analysis are  systematic differences, due
to variable simulation bos sizes and respective resolutions in
calculation of correlation functions. 

Basic data for our analysis are particle samples and density fields,
determined by the distribution of particles. To check our simulations
for possible errors we analysed in Section 3 density fields of
simulations for various box sizes, $L_0$, smoothing scales $R$, and
simulation epochs $z$. Results of this analyse are shown in
Fig.~\ref{fig:Fig1}. We see that the dispersion $\sigma$ varies with
box size $L_0$, smoothing scale $R$, and simulation epoch $z$ as
expected.  Thus variations in these parameters yield simulations with
good internal consistency.  Thus the influence of largest modes of
density perturbations is small.

Essential results of our analysis are presented in Tables \ref{Tab2}
-- \ref{Tab4}, and in lower panels of Fig.~\ref{fig:Fig6}.  This
Figure shows, first of all, that bias curves for increasing particle
density selection level $\rho_0$ form very regular sequences. This
suggests that errors in reducing bias parameters to identical
fractions of particles cannot be large.  However, there exist 
 differences in bias parameters, as found for different
box sizes -- at all evolutionary epochs $b$ values for larger box
sizes are larger.  These differences characterise the possible range
of uncertainty  in bias parameters.  We discuss this effect in more detail
in the next Section.

\section{Discussion}

Here we compare our data with results by other studies. Thereafter
we discuss the influence of a low-density homogeneous population in
voids. The Section ends with the discussion of our results for the
bias parameter.

\subsection{Comparison with earlier studies}

We begin the comparison with the analysis by 
\citet{Fry:1994vt}, who  derived galaxy correlation functions of samples of
increasing depth, using CfA, Southern Sky Redshift Survey (SSRS) and
IRAS redshift catalogs. Authors found that the correlation length $r$
of CfA samples increases with sample depth, $z_{lim} = 5000$~km/s to
$z_{lim} = 9000$~km/s, from $r=3.7~\Mpc$ to $r=5.8~\Mpc$ in real
space, and from $r=4.5~\Mpc$ to $r=6.8~\Mpc$ in redshift space.  A
similar growth is observed in SSRS and IRAS samples. These data
confirm results by \citet{Einasto:1986oh}, \citet{Einasto:1989vd} and
\citet{Einasto:1989cr} and \citet{Einasto:1994aa}, who explained this
increase as the approach to a representative sample of the universe
with increasing fraction of void volumes in samples, see the next subsection.

\citet{Fry:1996vj} investigated th evolution of bias, using a model in
which galaxies are formed at a fixed time and follow motions,
determined by the gravitational potential.  The bias parameter
decreases from $b=2.5$ at expansion factor $a=2$ to $b=1.15$ at
$a=20$. Fry concluded  that bias must have been larger in
the past.

\citet{Tegmark:1998yq} investigated the time evolution of bias using
perturbation theory, and adopting a time dependent galaxy formation
model.  Result for the bias parameter depend on the galaxy formation
model.  Between redshifts $z=5$ and $z=0$ the bias parameter decreases
in most models from $b \approx 2.5$ to $b \approx 1.2$. For all models
in a far future the bias parameter approaches to unity.

Now we compare our results with observational data on bias values for
high-redshift objects. \citet{Adelberger:1998ti} estimated correlation
functions for Lyman-break galaxies at redshift $z \sim 3$. Comparison
with models depends on cosmological parameters accepted, for
$\Omega_m=0.3$ and flat cosmogony authors obtained $b=4.0 \pm 0.7$.
This value is similar to our data for high luminosity (high particle
threshold limit $\rho_0$) samples. 

\citet{Song:2021tl} used Dark Energy Spectroscopy Instrument to study
galaxy clustering at redshift up to $z=1.6$ in several redshift
slices. Authors found that with increasing redshift from $z=0.5$ to
$z=1.1$ the linear bias parameter of Luminous Red Galaxies increases
from $b_1=2.22$ to $b_1=2.94$, and from $b_1=1.10$ to $b_1=1.45$ for
Emission Line Galaxies.

A study by \citet{Miyatake:2022aa} used CMB lensing
signals of 1.5 million galaxies at $z\sim 4$ to estimate
$\sigma_8$ and bias $b$ parameters. Result are  given in their Fig.~2 as
cosmological constraints for these parameters. For $\Omega_m=0.3$ the
lensing+clustering constraint suggests $\sigma_8(z=0) =0.5$ in good
agreement with our calculation, presented in Table~\ref{Tab1}. For
this $\sigma_8$ authors found bias parameter $b \approx 6$, similar 
to our samples with high particle density limit $\rho_0$.

\citet{Park:2022va} investigated the formation and morphology of the
first galaxies in the cosmic morning, $10 \ge z \ge 4$, using Horizon
Run cosmological simulations with gravity, hydrodynamics and various
astrophysics.  Among other results authors calculated correlation
functions of simulated galaxies in redshift range $z=5 - 7$, see
Fig.~11 by \citet{Park:2022va}.  Using data given in this Figure we
found bias parameter values $b=10,~7,~6$  for  redshifts $z=7,~6,~5$,
respectively.  These data by  \citet{Park:2022va} were found for
simulated galaxies with stellar  masses $M_\star \ge
2\times 10^9~M_\odot$.  Our Tables \ref{Tab2} -- \ref{Tab4} suggest
that these bias values are approximately equivalent to bias of our samples with
particle density limit $\rho_0 \approx 30$.

\subsection{Two populations of matter}

Differences in the distribution of matter and galaxies were noticed
already in the early stage of the study of cosmic web. Quantitatively
these differences can be described by differences of correlation
lengths in regions containing various amounts of voids in galaxy and
simulation samples.  In the present study we use $\Lambda$CDM
simulations and a wide range of cosmic epochs, starting from $z=30$,
and characterise biasing properties by the amplitude of the bias
functions -- ratios of correlation functions of galaxies and matter.
We use samples of particles with various local densities, which
characterise objects of different nature. As suggested by
\citet{White:1978} and \citet{Zeldovich:1982kl} and confirmed by
hydrodynamical models by \citet{Cen:1992kx, Cen:2000}, galaxies form
only in DM halos and not in under-dense regions in voids.  Thus the
matter is divided into two components, the clustered matter in
galaxies and clusters of galaxies, containing DM and visible galaxies,
and the unclustered matter in low density regions, consisting of DM
and rarefied baryonic gas, but no stars.

These two populations have very different spatial distribution.  The
clustered population with galaxies occupies small isolated regions,
the rest of the volume in voids is much larger, about 95\% of the
whole volume of the SDSS sample, for details see Fig.~2 and discussion in 
\citet{Einasto:2018aa}.  The amplitude of the
relative correlation function, measured by the bias parameter $b$, is
sensitive to the fraction of matter in the clustered population and in
the unclustered matter in voids, for a discussion see the next Subsection.  

Due to differences in spatial distribution, clustered and unclustered
populations have different influence to the biasing parameter $b$.
The unclustered matter in low density regions rises the bias parameter
from $b=1$ for the whole matter to $b \approx 1.5$ at the present
epoch, which corresponds to particle density limit $\rho_0 \approx 3$.
As shown by \citet{Einasto:2019aa}, this limit separates the
unclustered and clustered populations at the level of faintest
galaxies.  Further increase of bias parameter is due to the inclusion
to the sample brighter galaxies.

\subsection{The influence of a homogeneous population}

The dependence of the bias parameter on the fraction of
particles in the clustered population was suggested by
\citet{Saar:1983}, and studied in more detail by \citet{Einasto:1986oh},
\citet{Einasto:1987kw}, \citet{Gramann:1992ab}, \citet{Einasto:1994aa} and
\citet{Einasto:1999ku}.  This factor is crucial to understand our
results, thus we give here two  simple toy  models  to explain the idea.

\begin{figure}
\centering 
\hspace{1mm}  
\resizebox{0.35\textwidth}{!}{\includegraphics*{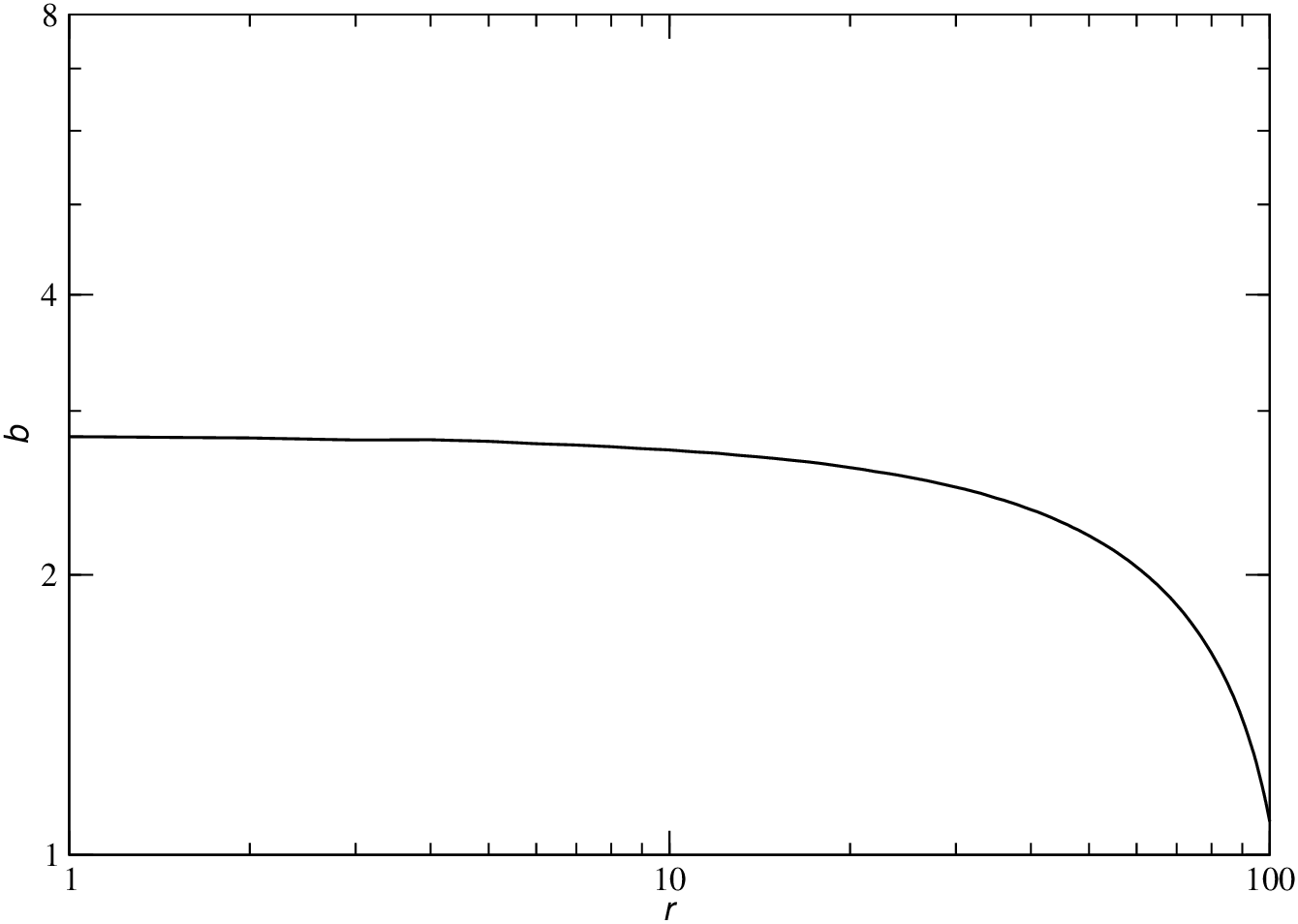}}
\caption{Bias function, found from the ratio of correlation functions $1+\xi(r)=RR_0(r)/RR_1(r)$.  }
\label{fig:Fig8} 
\end{figure}

The natural
estimator to determine the two-point spatial correlation function is
\be
1+\xi(r) = {DD(r) \over RR(r)},
\label{eq11}
\ee
where $DD(r)$ and $RR(r)$ are normalised counts of galaxy-galaxy and
random-random pairs at separation $r$.  Consider a volume of size
$V_0$, containing galaxies and systems of galaxies like supercluster
central regions. Denote counts of galaxy-galaxy and random-random
pairs as $DD_0(r)$ and $RR_0(r)$.  Now surround this volume with empty
space with no galaxies, and denote the total volume of this sample as
$V_1$. Galaxy-galaxy counts in the new volume are identical to counts
in the original volume, $DD_1(r) = DD_0(r)$. Random-random counts at
separation $r$, $RR_1(r)$ are, however lower, since the random sample
is diluted over a larger volume $V_1$.  This rises the amplitude of
the correlation function $1+\xi(r)$, the rise is proportional to the
ratio of volumes, $V_0/V_1$. To illustrate this effect we calculated
correlation functions for two samples, one for a sample with galaxies
of size $100~\Mpc$, and the other where this sample is located in a
cube of size $200~\Mpc$, containing outside the inner cube no
galaxies.  The bias function is given by the ratio of correlation
functions of both samples,  
$1+\xi(r)=RR_0(r)/RR_1(r)$.  This bias function is shown in
Fig.~\ref{fig:Fig8}. As we see, over most of the $r$ range it is
proportional to square root of the ratio of volumes, $V_1/V_0$.

The second model was suggested  by    \citet{Einasto:1999ku}, we give
here a summary of the discussion.

\begin{quotation}
Consider an idealized density field, which consists of a fluctuating
clustered component and a background of constant density, so that 
\be
D_m({\bf x}) = D_g({\bf x}) + D_s({\bf x});
\label{5.4}
\ee
here subscript $m$ is  for all matter, $g$ is for  galaxies (the
clustered component), and 
$s$ for the smooth component. 
The density contrast of the matter is 
\be
\delta_{m}= {D_{m} - \overline{D}_{m} \over \overline{D}_{m}};
\ee
or, applying (\ref{5.4}),   
\be
\delta_{m}= {D_g + D_s -(\overline{D}_g +
\overline{D}_s)
  \over \overline{D}_g + \overline{D}_s }.
\ee
Since  $D_s=\overline{D}_s$, we get 
\be
\delta_{m}= {D_g -\overline{D}_g \over \overline{D}_m} =
\delta_{g} {\overline{D}_g \over \overline{D}_m}.
\ee
In the last equation $\overline{D}_g/\overline{D}_{m}$ is the
fraction of matter in the clustered population, $F_c$; and we get
\be
\delta_m = \delta_g F_c.
\label{5.5}
\ee
According to traditional definition, Eq.~(\ref{bias0}),
$\delta_g/\delta_m =b$, and we get for the bias factor
\be
b={1 \over F_c}.
\label{5.6}
\ee

Equations (\ref{5.5}) and (\ref{5.6}) show that the subtraction of a
homogeneous population from the whole matter population increases the
amplitude of the correlation function (power spectrum) of the
remaining clustered population.  In this approximation biasing is
linear and does not depend on scale.  These equations have a simple
interpretation.  If we subtract from the density field a constant
density background but otherwise preserve density fluctuations, then
amplitudes of {\em absolute} density fluctuations remain the same, but
amplitudes of {\em relative} fluctuations with respect to the mean
density increase by a factor which is determined by the ratio of mean
densities.
\end{quotation}

Calculations made in the present
analysis show that the approximate equation (\ref{5.6}) is valid for
low particle density threshold $\rho_0 \le 1$. For larger $\rho_0$
values equation (\ref{5.6}) predicts too high values for the bias
parameter $b$. The reason for this deviation is the difference of the
density distribution in voids, which cannot be considered as constant,
as assumed in the Eq. ~(\ref{5.6}).

In this way we see  that the amplitude of the correlation function
measures the emptiness of samples of galaxies, quantified  by the fraction
of particles in voids and in the clustered population.

To check how accurately the relation (\ref{5.6}) helds we calculated
ratios $1/F_c$ for the present epoch $z=0$ for all $\rho_0$ values,
results are given in Tables~\ref{Tab2} to \ref{Tab4}. We see that in
all simulations bias parameter values for particle density threshold
$\rho_0 \le 3$ are rather close to values, found from the relation
(\ref{5.6}). This is expected, since bias functions $b(r)$ are almost
constant for $\rho_0 \le 3$, see Fig.~\ref{fig:Fig5}, i.e. correlation
functions have a similar shape and differ only by the amplitude, which
defines the bias parameter. For higher $\rho_0$ values differences in
shapes of correlation functions influence amplitudes of bias
functions, and the relation (\ref{5.6}) is not valid.

\subsection{Evolution of the bias parameter}

Basic results of our study are presented in Tables \ref{Tab2},
\ref{Tab3} and \ref{Tab4}, and in Figs.~\ref{fig:Fig5} and
\ref{fig:Fig6}.  The essential impression from Fig.~\ref{fig:Fig5} is
that bias is not a constant, as assumed in the classical theory (see
Eq.~(\ref{bias0})), but a function of the separation, $b(r)$.  The
function has two regions -- at small separations it describes the
distribution of particles (galaxies) in halos, at large separations it
describes the spatial distribution of halos (systems of galaxies).
The transition between these regions is determined by the scale of
halos (groups and clusters of galaxies), which depends on the
evolutionary epoch.  Halos are essentially virialised systems with
approximately constant physical sizes.  Our calculations are done in
comoving coordinates, thus in comoving coordinates halos were larger
in the past, as shown independently by \citet{Chiang:2013tq}

Fig. ~\ref{fig:Fig5} demonstrates that for all cosmic epochs bias
functions $b(r)$ form regular sequences, depending on galaxy
luminosity (particle density limit $\rho_0$). With increasing $\rho_0$
bias functions $b(r,\rho_0)$ increase. This is the dependence,
first described by \citet{Kaiser:1984}.  Now we have this relationship
for a large range of cosmic epochs $z$.

On scales not influenced by halos bias functions have approximately
constant amplitudes, which define bias parameters.  
Fig.~\ref{fig:Fig6} shows that  bias  parameter values $b(z)$ as
functions of the cosmic epoch, $b(z)$. 
Our data show that the bias parameter $b$ was higher in earlier cosmic
epochs.  In this way our analysis confirms earlier results, discussed
above, but in a larger range of  evolutionary epochs of the universe.
Physical reason for the increase of $b$ with epoch $z$ is simple -- in
earlier epochs the fraction of matter in voids was larger.  During the
evolution diffuse matter (DM and rarefied baryonic gas) flows out of
voids, as shown among others by \citet{Courtois:2017uv} and
\citet{Rizzi:2017wc}.

Data presented in lower panels of Fig. ~\ref{fig:Fig6} show that bias
parameter $b$ values for different particle density thresholds
$\rho_0$ depend on simulation box sizes: bias parameters $b$ are
highest for the simulation L1024, and lowest for the L256 simulation.
The reason for the dependence on simulation box size is not clear.  In
all cases correlation functions were calculated using particles
without additional smoothing, and bias parameters were reduced to
identical fractions of particles in the clustered population.  

As discussed above, low $\rho_0$ levels correspond to particles in
low-density regions and voids. The particle density threshold $\rho_0$
level which corresponds to faintest galaxies is known only
approximately. A better defined limit is the one, corresponding to
$L_\star$ and  Luminous Red Giant galaxies. As shown by
\citet{Einasto:2019aa} for the L512 simulation and the present epoch,
$L\ge L_\star$ galaxies have the same spatial distribution than DM
particles with density threshold $\rho_0 \ge 10$ in mean density
units,  see Fig.~10 by \citet{Einasto:2019aa}.  Correlation and bias
functions for this level are drawn in Figs.~\ref{fig:Fig5} and
\ref{fig:Fig6} with dashed lines.

As shown in \ref{fig:Fig6} and given in Tables \ref{Tab2}, \ref{Tab3}
and \ref{Tab4}, reduced bias parameters $b(10)$ are different in
simulations of various size.   To find
connections for L256 and L1024 simulations, a similar approach can be
applied, but this is outside the scope of the present study.

One reason for differences in bias parameters for different epochs is
the constant level of the fraction of particles in high-density
regions, used in the determination of the bias parameter. Actually groups
and clusters of galaxies grow with time by merging and infall of gas
and dark matter via filaments, which causes changes of fractions of
particles in high-density regions. To reduce our data to similar
populations of galaxy systems, we should take into account the growth
of systems. However, as shown by \citet{Park:2022va}, the growth of
masses (luminosities) of galaxies has little effect on the biasing
parameter. 

To conclude the discussion, we emphasise the role of different
populations in shaping bias functions and parameters. The unclustered
matter, consisting of DM particles and diffuse baryonic matter, and
filling about 95\% of the volume of the universe,  is
responsible in forming bias functions and parameters up to particle
threshold levels $\rho_0 \leq 3$.  Beyond this limit, $\rho_0 >3$,
particle density limited samples of DM simulations represent
luminosity limited samples of galaxies of various luminosity.  This is
the region initially discussed by \citet{Kaiser:1984}, and recently
studied by \citet{Norberg:2001aa}, \citet{Lahav:2002aa},
\citet{Verde:2002aa}, \citet{Tegmark:2004aa} and
\citet{Zehavi:2011aa}.  These studies describe well the relative
dependence of the bias parameter on the luminosity, in good agreement
with relative dependence of the bias parameter in our DM
simulations. However, these studies were not able to reduce bias
parameters of galaxies to that of the matter. The reason for this
difference could be  the insensitivity of methods, applied to find bias
parameters, to the smoothly distributed background of unclustered
matter. In contrast, the study by \citet{Park:2022va} allowed to take
into account both populations, and to find  bias parameters of simulated
galaxies in respect to matter.

\section{Conclusions}

In this paper we studied the evolution of the bias parameter using
numerical simulations of the evolution of the cosmic web. Our study is
based on three assumptions: (i) -- the $\Lambda$CDM model represents
real universe; (ii) -- particle density selected samples represent
galaxy samples; and (iii) -- sharp density threshold limit $\rho_0$
allows to select biased galaxy (particle) samples.  The novelty of our
approach lies in the use of numerical simulations in a large range of
evolutionary epochs, which allowed to take into account the influence
of both populations -- the smoothly distributed unclustered matter
with no visible galaxies, and the clustered matter with visible
galaxies.  We used several $\Lambda$CDM simulations and a wide range
of evolution epochs and particle density threshold levels to find bias
properties in a large range of cosmological parameter space.

Our basic results can be summarised as follows.

\begin{enumerate}

\item{} Bias is a function of particle separation $r$ and particle
  density selection level $\rho_0$, $b(r,\rho_0)$. On small
  separations, $r \le 10~\Mpc$, correlation and bias functions
  describe the distribution of particles (galaxies) in halos
  (clusters), on larger separations the distribution of halos (clusters).
  
\item{} For all cosmic epochs the bias parameter depends on two
  factors: the fraction of matter in the clustered population, and the
  particle density (galaxy luminosity)  limit of samples.  Gravity
  cannot evacuate voids completely, thus there is always some
  unclustered matter in voids, and the bias parameter of galaxies is
  always greater than unity, over the whole range of evolution epochs.
  
\item{} For all cosmic epochs bias parameter values form regular
  sequences, depending on galaxy luminosity (particle density limit),
  and decreasing with time.

\end{enumerate}

The present study allowed to find bias parameters in a much wider
parameter space in time and galaxy luminosity than made in earlier
studies. However, we consider the bias parameter for characteristic
luminosity $L_\star$ as a preliminary one, since simulations in cubes of
different size, L256 and L1024, give different results,
$b_\star = 1.5$ and $b_\star = 2.5$, respectively.  Do find a better
value of the characteristic bias parameter for the present epoch,
$b_\star$, and its evolution, a study is needed, which uses simulated
galaxies at various  epochs.

\section*{Acknowledgements}

We thank Gert H\"utsi for calculations of correlation functions and
stimulating discussions.  This work was supported by institutional
research funding IUT40-2 of the Estonian Ministry of Education and
Research, by the Estonian Research Council grant PRG803, and by
Mobilitas Plus grant MOBTT5. We acknowledge the support by the Centre
of Excellence``Dark side of the Universe'' (TK133) financed by the
European Union through the European Regional Development Fund.  The
study has also been supported by ICRAnet through a professorship for
Jaan Einasto.

\bibliographystyle{mnras} 

\bsp	
\label{lastpage}

\end{document}